\documentclass[12pt,a4paper]{report}

\usepackage[utf8]{inputenc}
\usepackage[T1]{fontenc}
\usepackage[english]{babel}
\usepackage{graphicx} 
\usepackage{amsmath, amssymb}
\usepackage{siunitx}
\usepackage{booktabs}
\usepackage{geometry}
\usepackage{hyperref}
\usepackage{setspace}
\usepackage{caption}
\usepackage{subcaption}
\usepackage{float}
\usepackage{enumitem}
\usepackage{titlesec}
\usepackage{pdflscape}
\usepackage{rotating}
\usepackage{tikz}
\hypersetup{
    colorlinks=true,
    linkcolor=black,
    citecolor=black,
    urlcolor=blue
}

\titleformat{\chapter}
  {\normalfont\huge\bfseries}
  {\thechapter.}
  {20pt}
  {}

\begin{document}

% Page de garde (non numérotée)
\pagenumbering{gobble}
\begin{titlepage}
\begin{center}

% ------------------ Logos en haut ------------------
\begin{minipage}{0.24\textwidth}
    \centering
    \includegraphics[width=\linewidth]{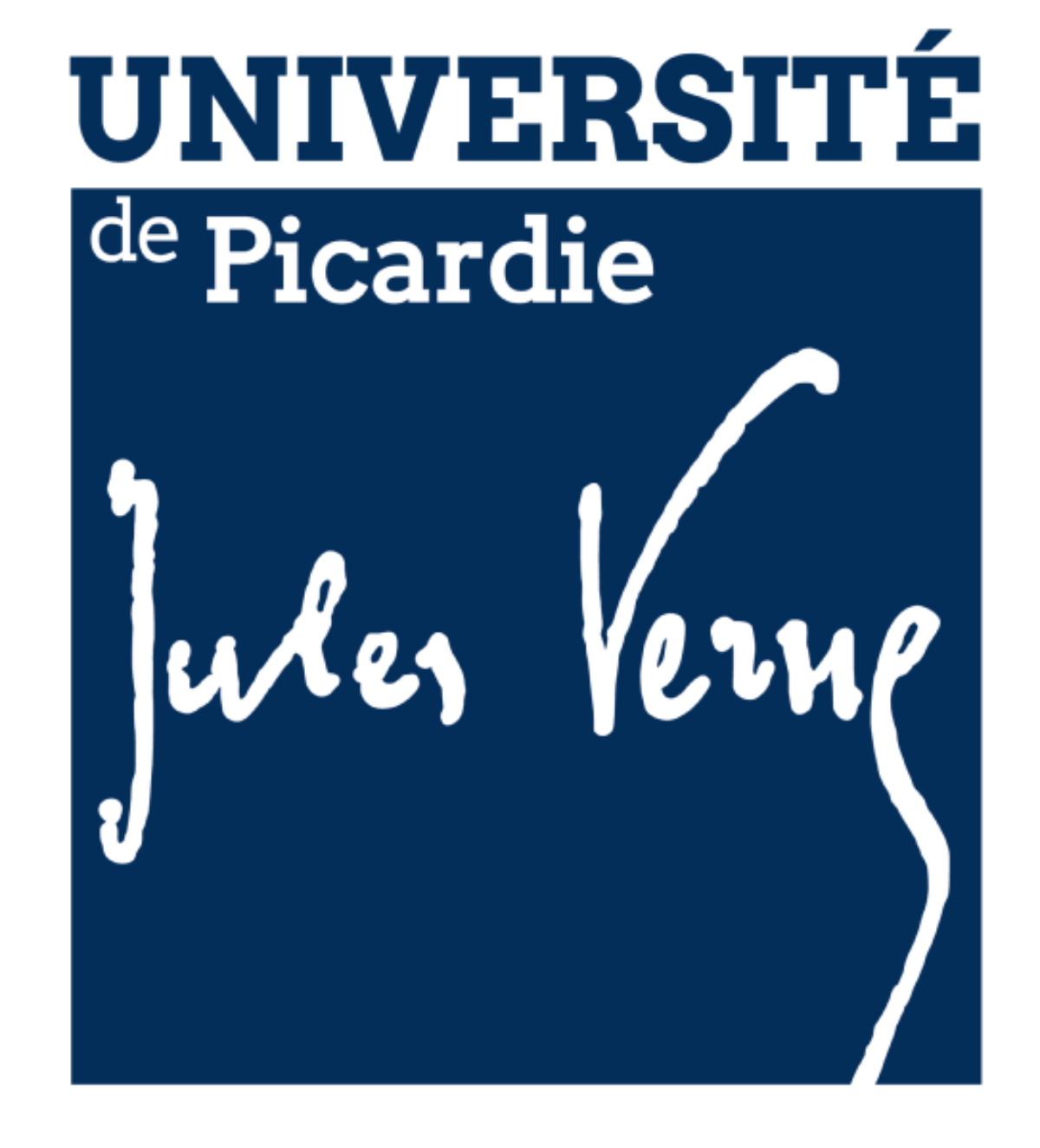}
\end{minipage}
\hfill
\begin{minipage}{0.24\textwidth}
    \centering
    \includegraphics[width=\linewidth]{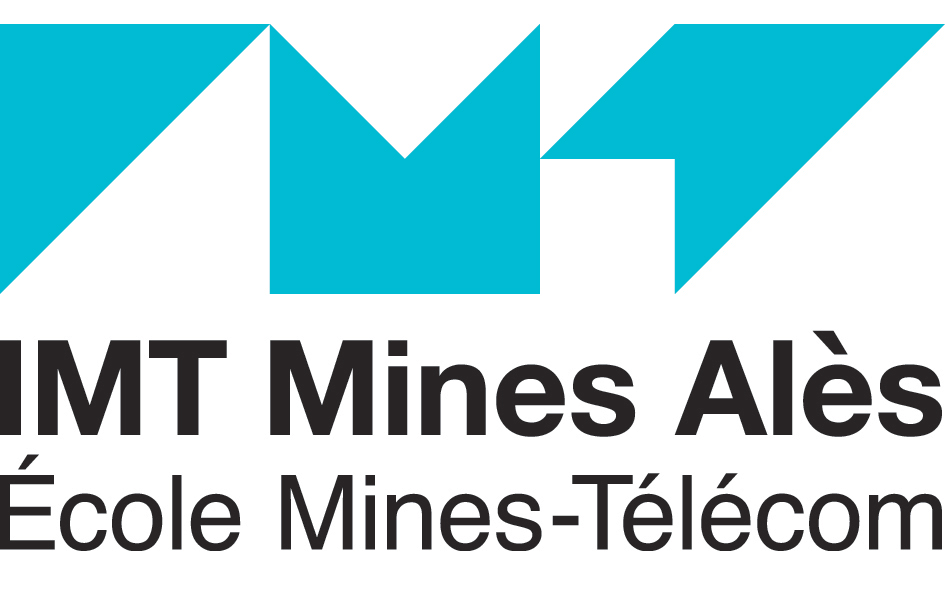}
\end{minipage}
\hfill
\begin{minipage}{0.24\textwidth}
    \centering
    \includegraphics[width=\linewidth]{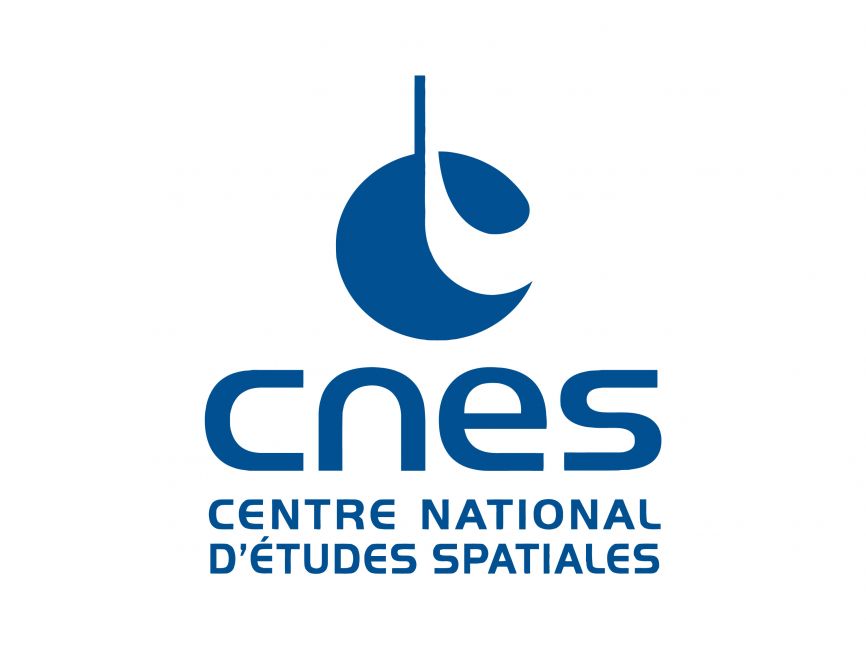}
\end{minipage}

\vspace{2cm}

% ------------------ Barre horizontale ------------------
\rule{\textwidth}{0.8pt}

\vspace{0.5cm}

% ------------------ Titre ------------------
{\Huge \textbf{MODÈFONE}}\\[0.4cm]
{\Large Modèle fantôme pour pression intracrânienne}\\[0.2cm]
{\small Phantom model for intracranial pressure}

\vspace{0.5cm}
CNES Parabole 2025
\vspace{0.5cm}

\rule{\textwidth}{0.8pt}

\vspace{2cm}

% ------------------ Auteurs ------------------
{\large
Célia \textsc{Batonon}\\
Heimiri \textsc{Monnier}\\
Arnaud \textsc{Gauberville}\\
Léna \textsc{Connesson}
}

\vspace{1cm}

% ------------------ Date ------------------
{\large \today}

\end{center}

% ------------------ Logo projet en bas à droite (ancré) ------------------
\begin{tikzpicture}[remember picture,overlay]
  \node[anchor=south east, xshift=-1cm, yshift=1cm] 
    at (current page.south east)
    {\includegraphics[width=0.5\textwidth]{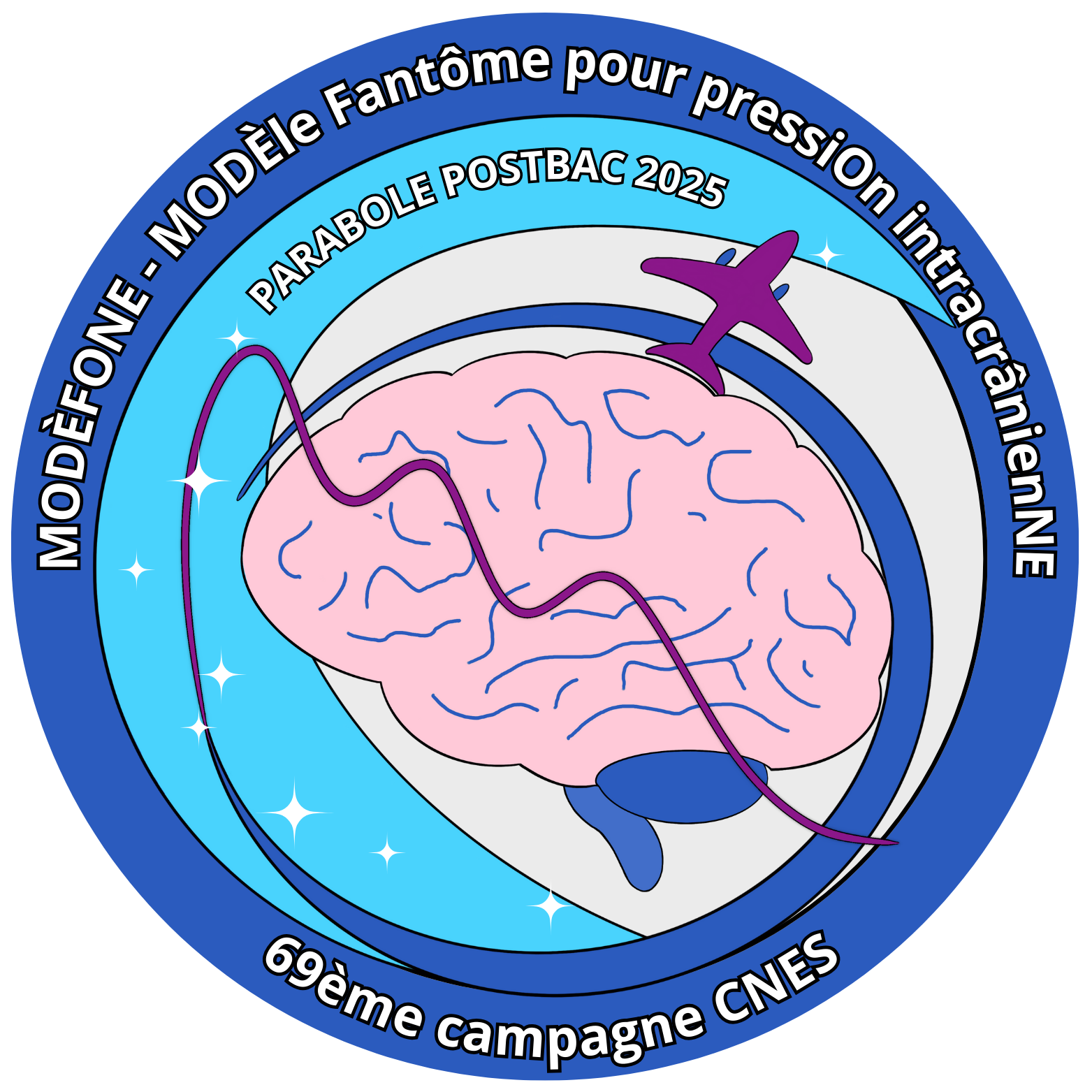}};
\end{tikzpicture}

\end{titlepage}

\cleardoublepage

% Pages liminaires
\pagenumbering{roman}
\chapter*{Abstract}

This report presents the MODÈFONE project, whose objective is to develop a simplified experimental model of the cerebrospinal system in order to investigate fluid–structure interactions and physiological adaptations under altered gravity conditions, with a particular focus on microgravity.

The experimental setup is based on a pulsatile hydraulic circuit reproducing systolic and diastolic dynamics, coupled with deformable elements simulating vascular compliance and a cranial compartment immersed in a fluid representing cerebrospinal fluid. This model enables the analysis of cranial and spinal pressures as well as their pulsatility.

The purpose of this report is to describe the design and the results of the experimental setup.

\vspace{1cm}

\noindent\textbf{Keywords:} cerebrospinal system, compliance, microgravity, phantom.

% Tables
\tableofcontents
\listoffigures
\chapter*{List of acronyms}
\addcontentsline{toc}{chapter}{List of acronyms}

\begin{tabular}{p{3cm} p{10cm}}
\textbf{CSF}  & Cerebrospinal Fluid \\
\textbf{ICP}  & Intracranial Pressure \\
\textbf{SANS} & Spaceflight Associated Neuro-ocular Syndrome \\
\textbf{MRI} & Magnetic Resonance Imaging \\
\textbf{BPM} & Beats Per Minute \\
\textbf{EF} & Ejection Fraction \\
\end{tabular}

\cleardoublepage

\cleardoublepage

% Corps du document
\pagenumbering{arabic}

% Chapitres
\chapter{Introduction}

\section{Scientific background}

The cerebrospinal system plays a central role in the mechanical protection of the brain and in the regulation of intracranial pressure (ICP). It relies on the dynamic interaction between three main compartments: the cerebral parenchyma, the vascular network, and the cerebrospinal fluid (CSF). These compartments are enclosed within a largely rigid cranial vault, extended by a more compliant spinal canal, which imposes strong mechanical constraints on intracranial volume variations.

At each cardiac cycle, the pulsatile inflow of arterial blood into the intracranial compartment induces a transient increase in cerebral blood volume. This volumetric variation is compensated by a displacement of CSF toward the subarachnoid spaces and the spinal canal, in order to limit ICP fluctuations and maintain intracranial homeostasis. This mechanism relies on the global compliance of the cerebrospinal system, defined as the ability of a compartment to deform in response to a pressure change. The relationship between intracranial volume and pressure variations directly depends on this compliance, which remains difficult to reliably measure in humans \cite{Goffin2017}.

\begin{figure}[H]
    \centering
    \includegraphics[width=0.85\textwidth, trim={0cm 0cm 0cm 0cm}, clip]{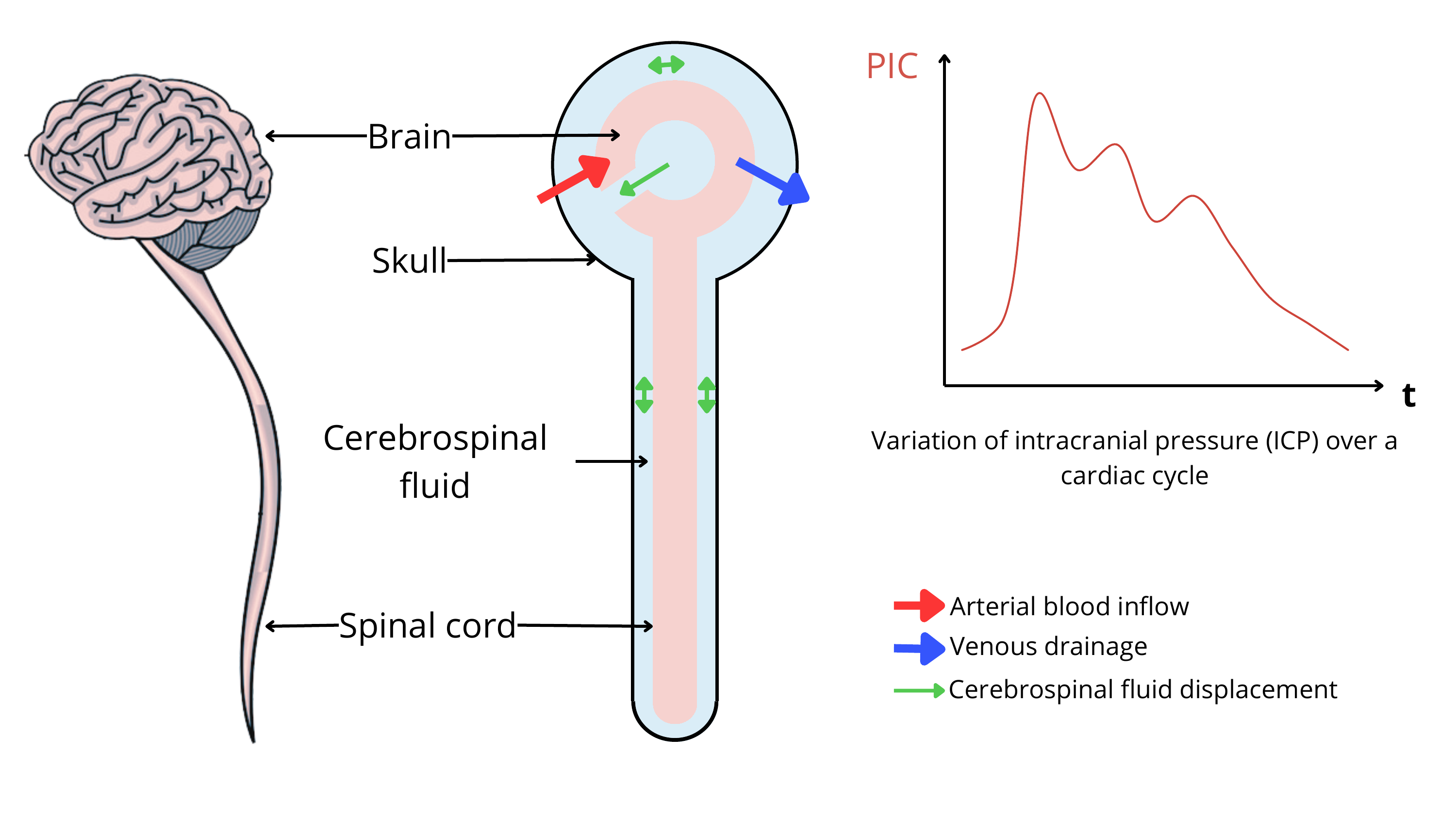}
    \caption{Simplified model of the cerebrospinal system.}
    \label{fig:schematic_cerebrospinal_system}
\end{figure}

ICP therefore constitutes a key parameter for studying cerebrospinal system dynamics. However, its direct measurement is invasive and requires the implantation of pressure sensors into the cerebral ventricles, which severely limits experimental investigations in healthy subjects \cite{Czosnyka2007}. Non-invasive approaches based on magnetic resonance imaging (MRI) have been developed, particularly using phase-contrast flow sequences that allow the measurement of pulsatile CSF and blood flow throughout the cardiac cycle \cite{Enzmann1993, Feinberg1984}. Nevertheless, these methods only provide indirect estimates of ICP and are generally restricted to supine positions.

On Earth, CSF dynamics and ICP are strongly influenced by gravity and body posture. Several studies have shown that cerebral venous drainage pathways depend on body position \cite{Gisolf2004, Alperin2005}. This venous redistribution directly affects intracranial volume variations and, consequently, ICP regulation.

A significant heterogeneity in venous drainage pathways has also been observed between individuals, even under normal physiological conditions \cite{Stoquart2009}. This interindividual variability complicates the global understanding of cerebrospinal dynamics and suggests that ICP regulation depends on multiple mechanisms, including anatomical, mechanical, and hemodynamic factors.

Under microgravity conditions, such as during spaceflight, the absence of gravity profoundly alters body fluid distribution. Venous return no longer benefits from the hydrostatic column and relies solely on cardiac suction during diastole. Studies conducted during parabolic flights and space missions have highlighted changes in jugular vein diameter and a redistribution of blood toward the upper body \cite{Lawley2017, Mari2023}. These findings suggest a major alteration of intracranial volume balance and CSF dynamics in microgravity.

These physiological disturbances are currently suspected to contribute to the development of \textit{Spaceflight Associated Neuro-ocular Syndrome} (SANS), a condition observed in some astronauts following long-duration missions. SANS is characterized by structural changes of the ocular globe, flattening of the posterior pole, and persistent visual impairments \cite{Alperin2017, Nelson2014}. Although the exact mechanisms underlying this syndrome remain debated, several hypotheses involve abnormal ICP variations, affecting either its mean value or its pulsatile component.

Recent studies even suggest that mean ICP in microgravity may be lower than values measured on Earth, challenging the initial hypothesis of chronic intracranial hypertension in astronauts \cite{Lawley2017}. These results highlight the complexity of interactions between gravity, venous circulation, CSF dynamics, and ICP regulation, and emphasize the need to develop new experimental tools to better understand these phenomena.

Thus, despite recent advances in imaging and space physiology, the fine dynamics of the cerebrospinal system in microgravity remain poorly understood. The difficulty of experimentally accessing ICP, combined with system complexity and spaceflight constraints, limits the ability to establish robust causal relationships between gravity, neurofluid circulation, and associated pathologies. This situation justifies the use of alternative experimental approaches capable of reproducing, in a controlled manner, the key mechanisms of cerebrospinal dynamics.

\section{Relevance of experimental models}

The study of ICP and CSF dynamics in humans currently relies on largely indirect or invasive approaches, which severely restrict experimental investigations, particularly in healthy subjects and in extreme environments such as microgravity.

Numerical models provide a complementary approach for investigating the theoretical mechanisms of cerebrospinal dynamics. They allow exploration of the influence of physiological parameters such as compliance, hydraulic resistances, and flow pulsatility. However, these models rely on the estimation of numerous parameters and critically require experimental reference data for validation, especially when extrapolating results to altered gravity conditions.

In this context, physical experimental models, referred to as \textit{phantoms} in biomedical engineering, represent a particularly relevant intermediate tool. These devices aim to reproducibly and controllably replicate fluid--structure interactions within the cerebrospinal system, while allowing direct instrumentation using pressure and flow sensors. They enable isolation of key physiological parameters, such as global system compliance or pressure pulsatility, and facilitate the study of their influence on ICP within a controlled environment.

Several cerebrospinal system phantoms have been developed over recent decades, primarily for studies conducted under terrestrial gravity conditions. Most of these models are designed as stationary test benches, intended for use in supine positions, and do not explicitly account for gravity effects on fluid distribution and venous drainage. While these approaches have advanced the understanding of CSF hydrodynamics, they remain limited when addressing the specific impact of gravity or its variations on intracranial dynamics.

Among the few experimental models that have incorporated gravity as a study parameter, the cerebrospinal system phantom developed at the CHIMERE UR 7516 laboratory represents an important reference \cite{Bouzerar2012}. This device enabled investigation of volume exchanges between the blood compartment and CSF while accounting for gravitational effects (supine and upright positions) on ICP. Using this model, it was demonstrated that a simplified physical reproduction of the cerebrospinal system is feasible.

However, this model presents certain limitations. In particular, the measured ICP is quasi-static (non-pulsatile), without faithful reproduction of the physiological pulsatility associated with the cardiac cycle. Yet, the pulsatile component of ICP plays a central role in cerebrospinal system dynamics and volumetric compensation mechanisms.

Finally, very few existing experimental models are compatible with platforms enabling access to variable gravity conditions, such as parabolic flights. These platforms nevertheless offer a unique opportunity to study CSF dynamics in real microgravity, in a repeatable and controlled manner.

Thus, physical experimental models appear as a promising alternative to bridge the gap between clinical observations, numerical modeling, and experimentation under altered gravity. By integrating physiological pulsatility, structural compliance, and gravity variations, these models provide a complementary approach to better understand the mechanisms governing ICP, particularly in the context of SANS.

\section{MODÈFONE project framework}

The MODÈFONE project (\textit{MODÈle Fantôme pour pressiOn intracrânienNE}) aims to design and instrument a simplified physical model of the cerebrospinal system capable of reproducing intracranial pressure variations induced by pulsatile blood flow.

The model is based on a rigid and deformable craniospinal structure filled with a fluid simulating CSF and coupled to a pulsatile hydraulic circuit mimicking cerebral blood inflow at each cardiac beat. Pressure and acceleration sensors allow real-time measurement of the system’s response to flow variations and changes in gravitational conditions.

\textbf{The primary objective of MODÈFONE is to investigate intracranial pressure evolution in real microgravity during parabolic flights and to compare these results with reference clinical data.} This project represents a first step toward the development of more comprehensive experimental models that will ultimately integrate ocular compartments, with potential applications in space physiology and biomedical research.

\chapter{Materials and methods}

\section{Experimental setup}

The experimental setup is designed as an integrated system combining the physical phantom structure, a hydraulic circuit, and an instrumentation chain enabling synchronized data acquisition (Figure~\ref{fig:fantome_1} and Figure~\ref{fig:fantome_2}).

\subsection{Phantom structure and mounting system}

The phantom provides a simplified representation of the craniospinal compartment. It consists of a rigid cranial part with spherical geometry connected to a cylindrical spinal part. The rigid components of the phantom were manufactured using 3D printing with ecoPETG filament (Forshape, Polyfab3D), selected for its good mechanical strength, dimensional stability, and compatibility with confined environments. All components were assembled in a watertight manner (using seals and varnishing of the 3D-printed parts with chloroform) to allow fluid circulation between compartments.

The cranial compartment has an internal diameter of approximately 13~cm, while the spinal compartment is approximately 13~cm in length with an internal diameter of about 2.8~cm. The connection between the two compartments is ensured by a rigid conduit, allowing volume exchange and transmission of pressure variations between the cranial and spinal compartments.

A deformable section is integrated into the spinal part of the phantom. It consists of a latex material, specifically a male condom, selected for its high compliance and its ability to exhibit deformable behavior under small pressure variations. This deformable section introduces an effective compliance into the system and promotes fluid displacement in response to pressure changes.

The phantom is entirely filled with water to simulate cerebrospinal fluid (CSF). The total fluid volume contained in the system is approximately 1.4~L, distributed between the cranial and spinal compartments according to their respective geometries. This volume is consistent with the order of magnitude of physiological human CSF volume and allows measurable pressure variations at the scale of the model.

The phantom design follows a proportional approach rather than a strict anatomical scaling. Flow resistances between the cranial and spinal compartments were determined based on orders of magnitude reported in the literature, while maintaining resistance ratios compatible with human physiology. The objective is not to faithfully reproduce human anatomy, but to preserve the relative mechanisms of volume and pressure exchange between compartments.

The phantom is mounted on a rigid mechanical support fixed to an aluminum plate integrated into a sealed rack (Figure~\ref{fig:fantome_2}). The mounting system allows adjustment of the phantom orientation relative to the vertical, with angles ranging from 0° to 90°. This modularity enables investigation of posture-related effects on fluid dynamics and intracranial pressure (ICP). A schematic of the complete setup is provided to illustrate the phantom structure.

\begin{figure}[H]
    \centering
    \includegraphics[width=0.9\textwidth]{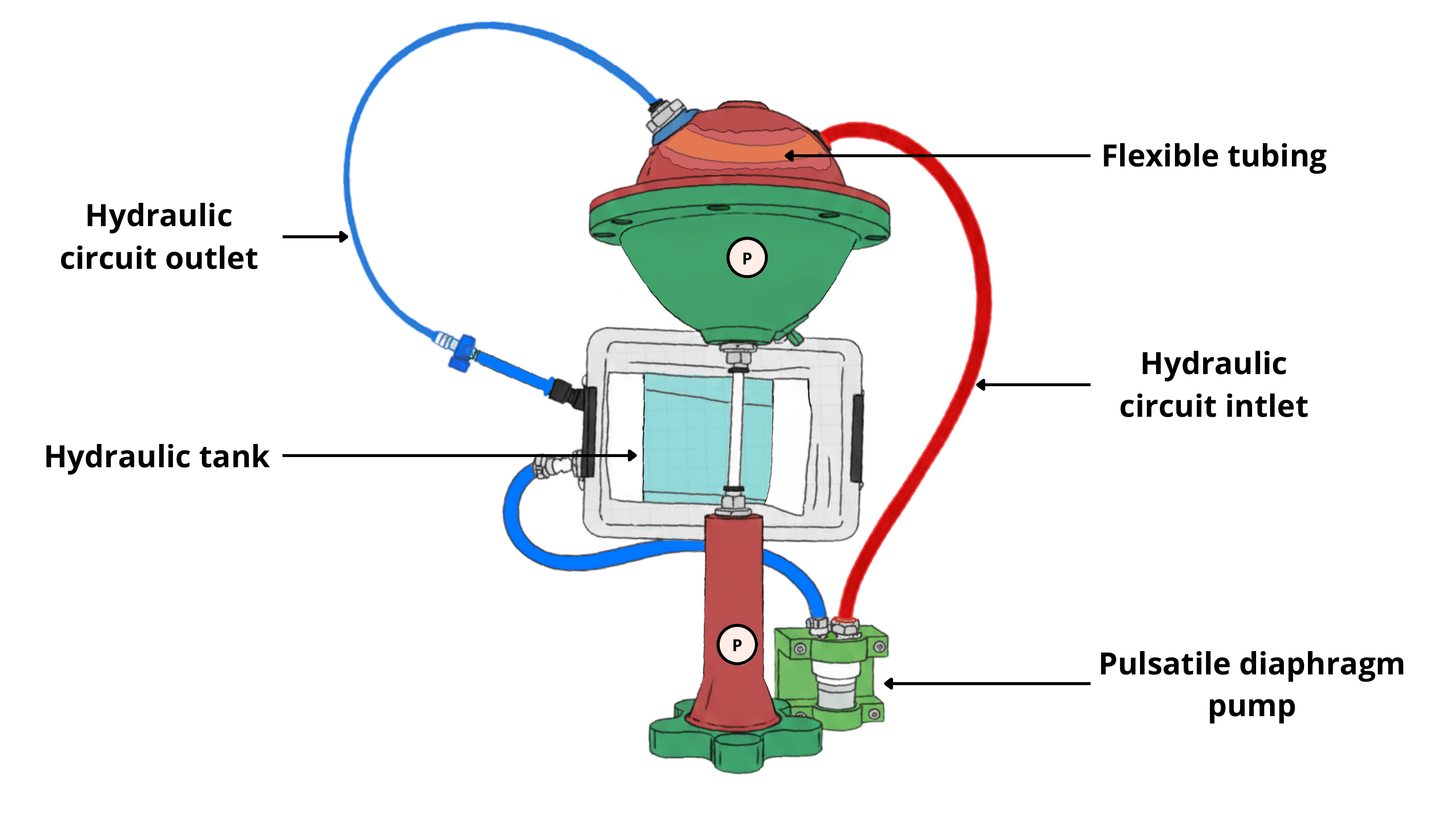}
    \caption{Schematic representation of the MODÈFONE phantom.
    \\
    \small P: pressure sensor.}
    \label{fig:schematic_MODEFONE}
\end{figure}

\subsection{Pulsatile hydraulic system}

The phantom structure is coupled to a closed hydraulic circuit designed to generate a pulsatile excitation of the system. This circuit aims to reproduce, in a simplified manner, the volume variations induced by the pulsatile inflow of blood into the intracranial compartment, which are responsible for fluid displacement and ICP fluctuations.

The hydraulic system is based on a 12~V electronically controlled diaphragm pump capable of delivering a controlled oscillatory flow rate. The pump is driven to impose a periodic excitation of the circuit, with heart rate and ejection fraction selected within ranges compatible with human cardiac physiology. This approach is supported by recent studies demonstrating the relevance of using small pumps to generate representative pulsatile flow profiles in experimental biomedical systems \cite{OpenHeart2024}.

The maximum flow rate of the system is on the order of 1 to 1.2~L/min, enabling the generation of measurable pressure variations while remaining compatible with the mechanical constraints of the phantom and the operating ranges of the sensors.

A key section of the circuit consists of a deformable tube placed inside the cranial compartment of the phantom. Under the flow imposed by the pump, this tube undergoes cyclic expansion and contraction, inducing volume variations transmitted to the surrounding water. This mechanism reproduces a pulsatile volumetric excitation of the system, analogous to that induced by cerebral blood flow pulsatility, without directly imposing pressure variations.

The remainder of the circuit consists of rigid or semi-rigid tubing, ensuring stable flow conditions and limiting uncontrolled pressure losses. This configuration ensures that the dynamic response of the system is dominated by the deformable elements of the phantom and the circuit, while guaranteeing experimental reproducibility.

\subsection{Instrumentation, control, and data acquisition}

The experimental setup is instrumented to measure the physical quantities required for system dynamics analysis, while ensuring control of the pulsatile hydraulic circuit. All instrumentation is designed to operate reliably in a parabolic flight environment, with synchronized data acquisition.

Pressure variations within the phantom are measured using two silicon piezoresistive pressure sensors (MS5837-02BA, Senzooe), positioned respectively in the cranial and spinal compartments. These sensors provide a measurement range suitable for the low pressure variations expected in the system and are powered with a 3~V DC supply. They are connected to the electronic system via interface boards, ensuring reliable and watertight integration within the phantom structure.

The gravitational conditions experienced by the device during flight are recorded using a triaxial accelerometer (MMA8451Q, Adafruit). This sensor continuously measures accelerations and enables precise identification of normal gravity, hypergravity, and microgravity phases. Accelerometer data are used to synchronize pressure signals with the different flight phases.

Pump control and data acquisition are handled by an Adafruit ESP32 Feather V2 microcontroller. This microcontroller enables both control of the oscillatory flow rate via a DC Motor + Stepper FeatherWing (Adafruit) and acquisition of sensor signals. Pump control is achieved through modulation of the voltage applied to the motor, allowing fine adjustment of both the frequency and amplitude of the pulsatile excitation.

The entire electronic system is powered by dedicated AC/DC converters providing stabilized voltages suitable for the various components (3~V and 12~V). All electronic elements are housed in a sealed enclosure fixed to the experimental rack plate.

Sensor data are transmitted in real time to a laptop computer connected to the microcontroller via a USB link. Data are continuously recorded throughout the experiment, enabling post hoc analysis of ICP variations as a function of hydraulic parameters and gravitational conditions.

\begin{figure}[H]
    \centering
    \includegraphics[width=1.0\textwidth]{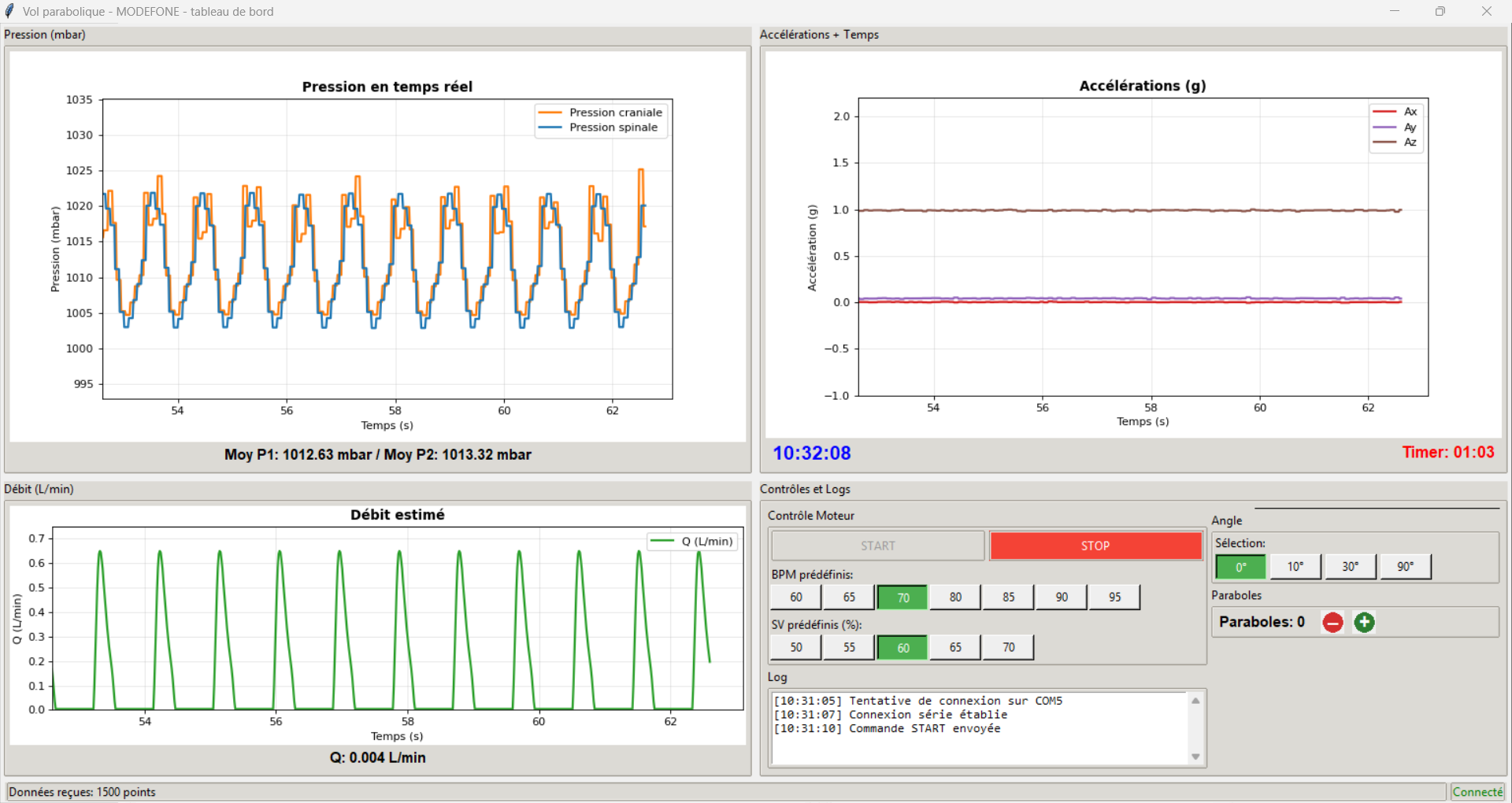}
    \caption{MODÈFONE experimental interface.}
    \label{fig:interface_MODEFONE}
\end{figure}

\section{Experimental platform: parabolic flights}

The experiment was conducted aboard the AirZero-G aircraft (Novespace), which performs parabolic flight maneuvers providing repeated and controlled access to variable gravity phases. This experimental platform is particularly suited for studying gravity-dependent physiological and biomechanical phenomena, offering real microgravity conditions over short but reproducible durations.

A parabolic flight consists of a sequence of parabolas, each comprising three distinct phases: a normal gravity phase (approximately 1g), followed by a hypergravity phase (approximately 1.8g), and then a microgravity phase close to 0g. The microgravity phase typically lasts between 20 and 22 seconds, during which gravitational effects are strongly attenuated. These phases are repeated 31 times during a single flight, allowing acquisition of comparable datasets under different gravitational conditions.

The experimental device is installed within a sealed rack fixed to the aircraft structure. This integration ensures the safety of both the system and the crew while maintaining mechanical stability of the phantom and associated equipment. The entire setup is designed to operate autonomously during flight phases. Experimental intervention is limited to adjusting motor parameters via the interface and modifying the phantom orientation by opening the rack.

\begin{figure}[H]
    \centering
    \includegraphics[width=0.8\textwidth]{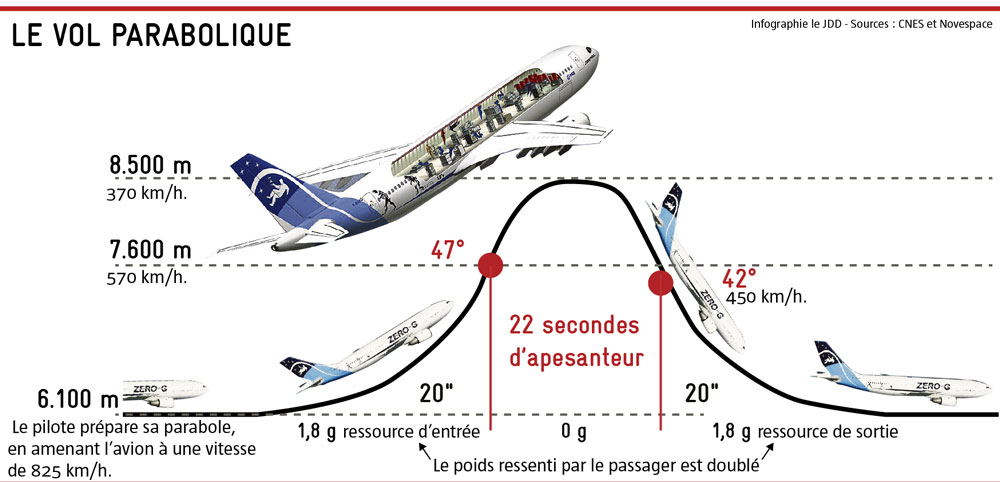}
    \caption{Parabolic maneuver during a parabolic flight.}
    \label{fig:manoeuvre_parabole}
\end{figure}

\section{Experimental protocol}

The experimental protocol implemented during the parabolic flight relies on controlled variation of pulsatile hydraulic system parameters and phantom orientation in order to study their influence on CSF dynamics and ICP under different gravitational conditions. The protocol was defined prior to the flight and structured into successive groups of parabolas, each corresponding to a specific experimental configuration.

\subsection{Experimental parameters and variables of interest}

The experiment was designed to investigate the influence of two main factors on ICP dynamics: gravity level and phantom orientation relative to the gravity vector.

Gravity level was explored during the different flight phases, including normal gravity (1g), hypergravity (1.8g), and microgravity (0g). Phantom orientation was modified only during normal gravity phases, with three configurations tested: horizontal position (0°), intermediate position (30°), and vertical position (90°). Stabilization periods were respected after each orientation change before resuming measurements.

Pump parameters used to generate pulsatile flow in the hydraulic circuit were not considered as independent study variables. Heart rate (BPM) and ejection fraction (EF) were adjusted according to phantom orientation to reproduce physiologically coherent hemodynamic conditions. The selected values correspond to human orders of magnitude, with heart rates ranging from 70 to 83~BPM and EF values between 47 and 63\%.

Accordingly, in the remainder of this work, observed ICP variations are analyzed solely with respect to gravity level and phantom orientation, with motor parameters considered as control variables ensuring physiological consistency of experimental conditions.

\subsection{Data acquisition during parabolas}

For each parabola, pressure data, estimated flow rate derived from pump control voltage, and acceleration data were continuously recorded. Normal gravity, hypergravity, and microgravity phases were identified a posteriori using the onboard accelerometer signal. Pressure signals were then analyzed as a function of gravitational phases and imposed experimental parameters.

Repetition of configurations across multiple parabolas improves analysis robustness and allows comparison of system responses under different gravitational conditions and phantom orientations for a given hydraulic excitation.

\section{Data processing}

Data acquired during parabolic flights were post-processed to extract variables of interest and enable comparative analysis between experimental conditions. Processing included estimation of hydraulic flow rate, automatic segmentation of flight phases and simulated cardiac cycles, and filtering of pressure signals.

\subsection{Pump flow rate estimation}

As the experimental setup does not include a flowmeter in the hydraulic circuit, pump flow rate was estimated indirectly from motor control parameters. The maximum pump flow rate is known for a given maximum supply voltage. A proportional relationship between applied motor voltage and generated flow rate was established, enabling estimation of instantaneous pump flow rate from the control voltage.

This approach assumes quasi-linear pump behavior within the operating range used during the experiment. Although it does not provide absolute flow measurements, it is sufficient for relative comparison of experimental conditions and analysis of pressure pulsatility effects.

\subsection{Segmentation of flight phases and cardiac cycles}

Flight phases (normal gravity, hypergravity, and microgravity) were automatically identified from the triaxial accelerometer signal. A detection algorithm segments data based on measured acceleration levels, ensuring reproducible identification of each parabola phase.

Similarly, simulated cardiac cycles were detected from the pump control signal. This automatic detection enables segmentation of data into successive cycles and analysis of system response over a typical cardiac cycle, independently of variable gravity phase duration.

\subsection{Preprocessing and analysis of pressure signals}

Pressure signals measured in the cranial and spinal compartments constitute the primary variables of interest. To reduce measurement noise and facilitate analysis, data preprocessing was applied. Signals were first linearly interpolated to ensure temporal continuity, then smoothed using a Savitzky--Golay filter, which preserves rapid pulsatile variations while attenuating high-frequency noise.

Data analysis focused on two main indicators. The first corresponds to the mean pressure value measured in each compartment, computed separately for the different gravity phases. The second concerns the pulsatile pressure component, analyzed over a simulated cardiac cycle and compared across normal gravity, hypergravity, and microgravity phases.

\subsection{Statistical analysis}

Statistical comparisons between gravity conditions were performed using Welch’s analysis of variance (Welch ANOVA). This test was selected for its robustness to unequal sample sizes and potential heteroscedasticity, conditions encountered in parabolic flight experiments.

Welch ANOVA was applied to mean pressure values as well as to pulsatility indicators extracted for each gravity phase. When significant differences were detected, results were interpreted considering intra- and inter-parabola variability. This statistical approach allows evaluation of gravitational effects on ICP while limiting biases related to the experimental data structure.

\begin{figure}[H]
    \centering

    \begin{subfigure}[t]{0.7\textwidth}
        \centering
        \includegraphics[width=\linewidth]{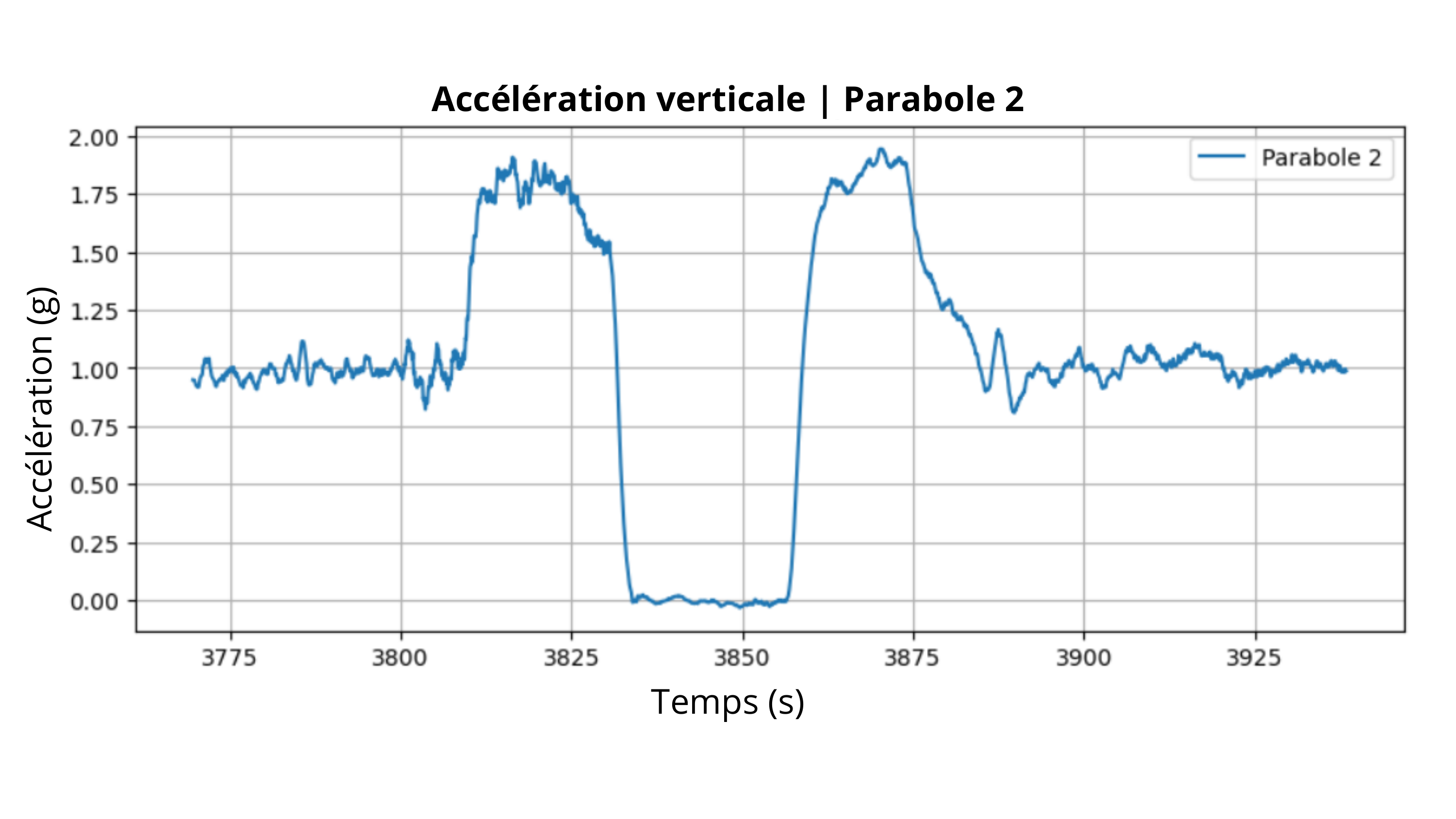}
        \caption{Detection of a parabola.}
        \label{fig:post_parabole_detection}
    \end{subfigure}

    \vspace{4mm}

    \begin{subfigure}[t]{0.7\textwidth}
        \centering
        \includegraphics[width=\linewidth]{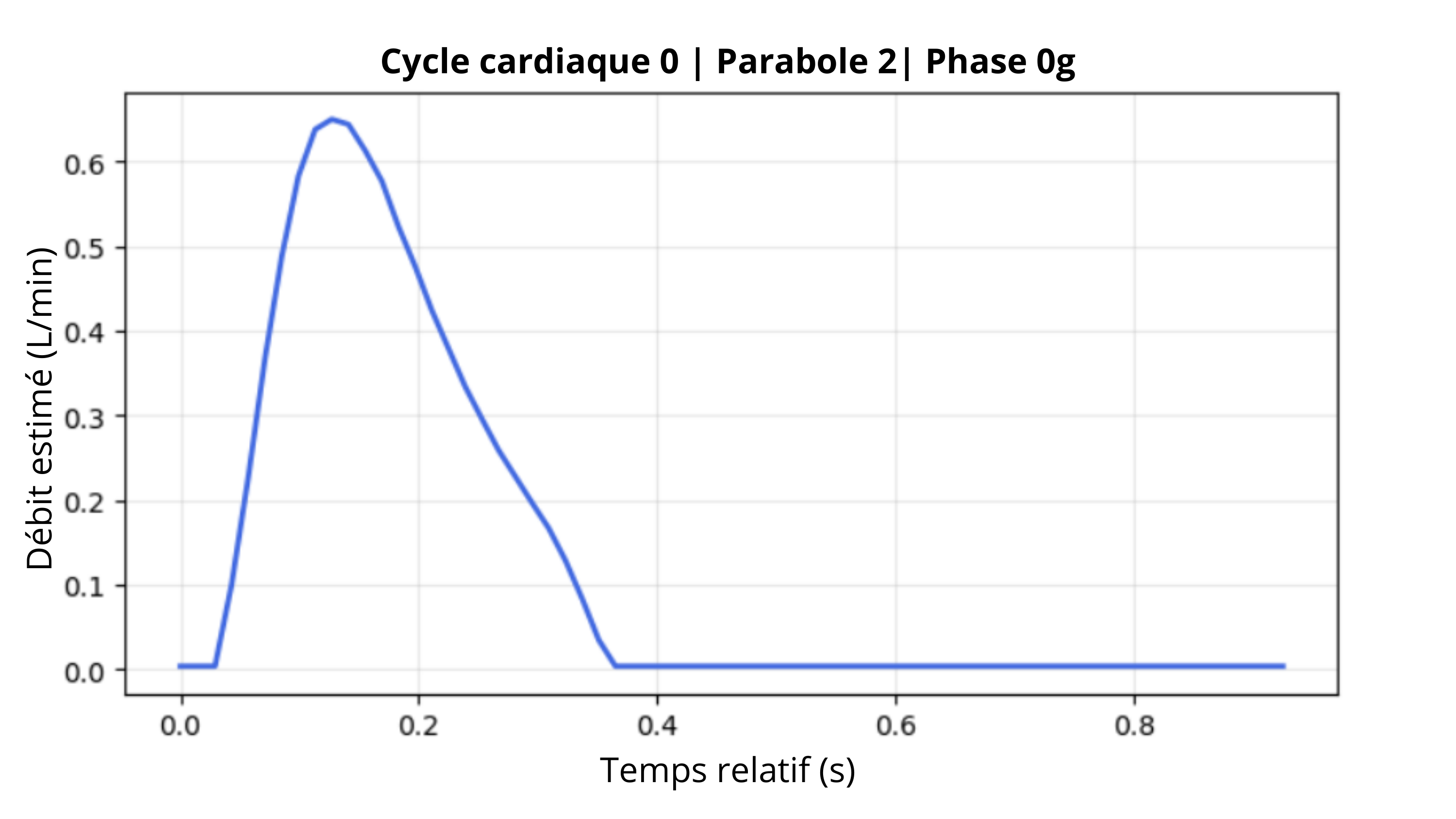}
        \caption{Detection of a cardiac cycle.}
        \label{fig:post_cycle_detection}
    \end{subfigure}

    \caption{Post-processing of experimental data.\\
    Débit estimé: Estimated flow; Accélération: Acceleration; Temps relatif: Relative time.}
    \label{fig:post_traitement}
\end{figure}

\chapter{Results}

\section{Effects of gravity level and orientation on mean pressure values}

\subsection{Mean intracranial pressure}

Mean intracranial pressure (ICP) in the phantom was analyzed to assess the effects of gravitational conditions and model orientation on the quasi-static component of pressure. The ICP considered here corresponds to the pressure measured in the intracranial compartment of the water-filled phantom and does not represent a direct physiological human pressure. Values are reported in hectopascals (hPa), as measurements were performed onboard the aircraft where atmospheric pressure is lower than at ground level.

Mean ICP values were extracted only for conditions in which pump settings were kept constant. For the 0° and 30° orientations, the pump was set to 70~BPM with an ejection fraction (EF) of 60\%, whereas for the vertical orientation (90°), parameters were set to 80~BPM and 50\%.

\begin{figure}[h]
    \centering
    \includegraphics[width=1.0\textwidth]{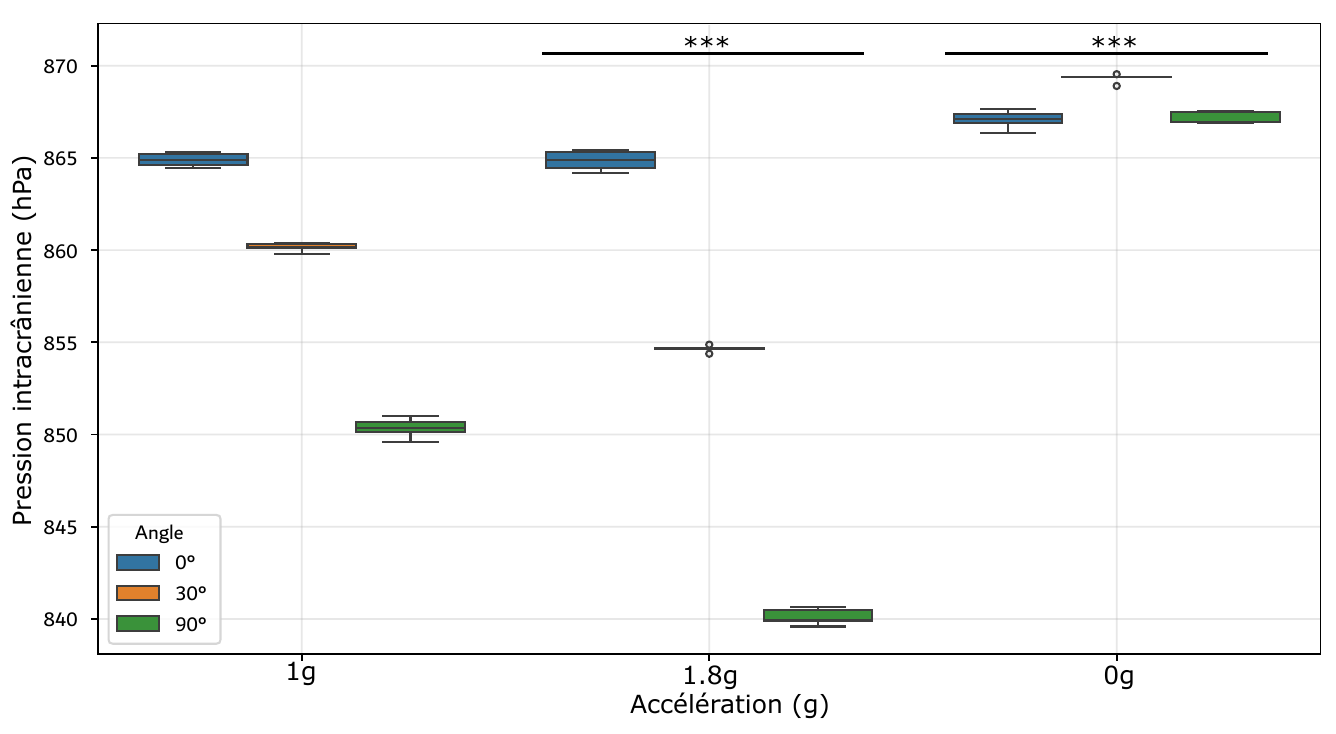}
    \caption{Mean intracranial pressure as a function of gravitational conditions (1g, 1.8g, and 0g) and phantom orientation (0°, 30°, and 90°).
    \\
    \small \textit{***}: $p < 0.001$.}
    \label{fig:pic_mean}
\end{figure}

Figure~\ref{fig:pic_mean} shows mean ICP values for the different gravity phases (1g, 1.8g, and 0g) and the three phantom orientations. Under normal gravity (1g), mean ICP was 865~hPa in the horizontal position (0°), 860~hPa at 30° inclination, and 850~hPa in the vertical position (90°). These results indicate a progressive decrease in mean ICP as the phantom orientation shifts from horizontal to vertical.

Under hypergravity (1.8g), mean ICP measured in the horizontal position (0°) was 864.7~hPa, very close to the value observed at 1g. For 30° and 90° orientations, mean values were 854.6~hPa and 840~hPa, respectively. The decrease in mean ICP with increasing phantom angle is therefore also observed under hypergravity, with a larger magnitude than under normal gravity.

Under microgravity (0g), mean ICP measured in the horizontal position (0°) was 867~hPa. Comparable values were observed for the other orientations, with a mean ICP of 869~hPa at 30° and 867~hPa in the vertical position (90°). In contrast to normal gravity and hypergravity conditions, mean ICP does not exhibit a monotonic variation with orientation under microgravity, suggesting a different system response in reduced-gravity conditions.

Statistical analyses confirm these observations, with a significant effect of gravity on mean ICP (Welch ANOVA, $p < 10^{-10}$) and a significant effect of phantom orientation (Welch ANOVA, $p < 10^{-12}$).

\subsection{Mean spinal pressure}

\begin{figure}[h]
    \centering
    \includegraphics[width=1.0\textwidth]{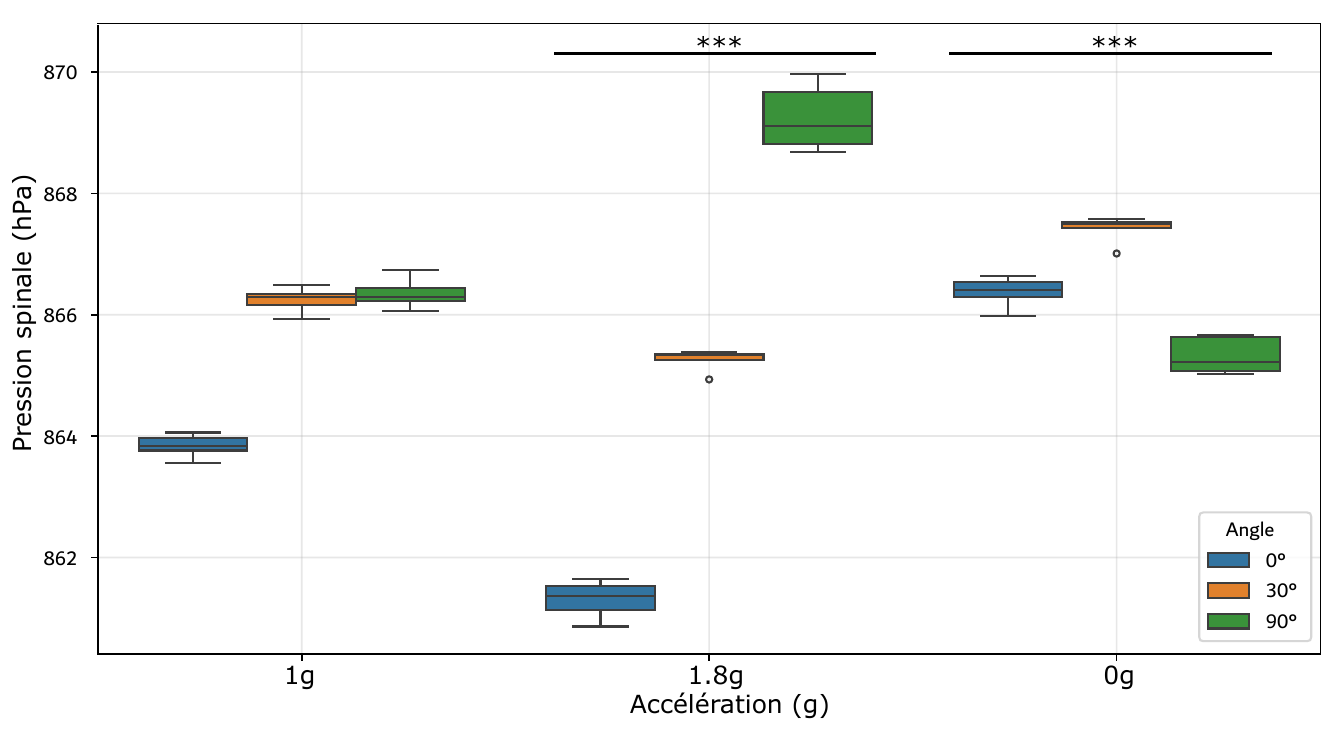}
    \caption{Mean spinal pressure in the phantom as a function of gravitational conditions (1g, 1.8g, and 0g) and phantom orientation (0°, 30°, and 90°).
    \\
    \small \textit{***}: $p < 0.001$.}
    \label{fig:p_spinal_mean}
\end{figure}

Similarly, Figure~\ref{fig:p_spinal_mean} shows mean spinal pressure values measured across the different gravity phases (1g, 1.8g, and 0g) and the three phantom orientations (0°, 30°, and 90°).

Under normal gravity (1g), mean spinal pressure was 864~hPa in the horizontal position (0°), 866~hPa at 30° inclination, and 866.3~hPa in the vertical position (90°). Unlike the intracranial compartment, spinal pressure shows a slight increase with phantom angle, although differences between positions remain limited.

Under hypergravity (1.8g), mean spinal pressure was 861.4~hPa in the horizontal position (0°), 865~hPa at 30° inclination, and reached 869~hPa in the vertical position (90°). A progressive increase in spinal pressure with orientation is observed in this condition, with a more pronounced difference than under normal gravity.

Under microgravity (0g), mean spinal pressure was 866~hPa in the horizontal position (0°), 867~hPa at 30° inclination, and 865~hPa in the vertical position (90°). Values remain globally close to each other, without a clear monotonic trend as a function of orientation.

Statistical analyses confirm these observations. A significant effect of gravity on mean spinal pressure was identified (Welch ANOVA, $p = 1.9 \times 10^{-5}$), although with a smaller magnitude than for ICP. The effect of phantom orientation is, however, very pronounced, with a significant influence of position on mean spinal pressure (Welch ANOVA, $p < 10^{-22}$).

Overall, mean pressure values measured in the intracranial and spinal compartments are of the same order of magnitude under normal gravity (1g) for the horizontal position (0°).

Under normal gravity (1g), a progressive decrease in mean ICP is observed as phantom orientation increases, with a difference of approximately 15~hPa between the horizontal position (865~hPa) and the vertical position (850~hPa). Spinal pressure shows a more moderate evolution, with a slight increase of about 2.3~hPa between 0° (864~hPa) and 90° (866.3~hPa). These results reflect a differentiated behavior between intracranial and spinal compartments as a function of posture.

Under hypergravity (1.8g), the same general trend is observed. Mean ICP decreases with orientation, from 864.7~hPa in the horizontal position to 840~hPa in the vertical position, i.e., a decrease of approximately 25~hPa. Conversely, spinal pressure increases markedly with phantom angle, with a difference of about 7.6~hPa between 0° (861.4~hPa) and 90° (869~hPa). This opposing trend between the two compartments is particularly pronounced under hypergravity.

Under microgravity (0g), the behavior of both compartments differs from that observed at 1g and 1.8g. Mean ICP values are slightly higher at the intermediate orientation, with 869~hPa at 30°, compared with 867~hPa at 0° and 90°. A similar trend is observed for spinal pressure, with a maximum value of 867~hPa at 30°, compared with 866~hPa at 0° and 865~hPa at 90°. Thus, under microgravity, intracranial and spinal pressures do not follow a monotonic evolution with orientation, but instead show a slight maximum at the intermediate phantom position.

\section{Effects of gravity level and orientation on pressure pulsatility}

\subsection{Pulsatility across gravity phases}

\paragraph{ICP pulsatility at 0°.}

At 0°, intracranial pressure (ICP) pulsatility is strongly modulated by the gravity phases. Waveforms averaged over a cardiac cycle show stable and reproducible dynamics across all analyzed cycles (Figure~\ref{fig:pulsatile_pic_0_90}), enabling reliable comparisons between gravitational conditions.

Under normal gravity, ICP exhibits a complex pulsatile morphology characterized by three distinct peaks over the cardiac cycle. Transition to hypergravity (1.8g) is associated with a modification of this structure, with attenuation of the first peak and an overall amplification of pressure variations. Conversely, under microgravity, pulsation amplitude decreases and returns to values comparable to those observed at 1g, while the first two peaks appear more pronounced. This suggests a redistribution of pulsatile dynamics rather than a simple reduction in amplitude.

This qualitative evolution is reflected in ICP amplitude values. At 1g, the mean amplitude is 13.9~hPa. It increases markedly under hypergravity, reaching 19~hPa, indicating enhanced pressure fluctuations over the cardiac cycle. Under microgravity, amplitude decreases to 12.3~hPa, returning to levels close to those measured under normal gravity (Figure~\ref{fig:pic_pulsatility_boxplot}).

In the horizontal orientation of the phantom, these results highlight a pronounced increase in ICP pulsatility under hypergravity, followed by attenuation under microgravity, suggesting a non-linear response of the intracranial system to gravity variations.

\paragraph{ICP pulsatility at 90°.}

In the vertical position (90°), ICP pulsatility remains strongly dependent on gravity conditions (Figure~\ref{fig:pulsatile_pic_0_90}), with overall amplitudes higher than those measured in the horizontal position. Under normal gravity and hypergravity, the waveform retains a well-defined three-peak structure over the cardiac cycle, with the second peak consistently dominant. This morphological similarity suggests that the vertical orientation promotes robust pulsatile dynamics that are only weakly altered when transitioning from 1g to 1.8g.

Under microgravity, ICP morphology is instead profoundly modified: the main peak becomes more rounded and the second peak is strongly attenuated, indicating altered pulsatile dynamics relative to terrestrial gravity conditions.

These qualitative differences are accompanied by marked amplitude changes. At 1g, amplitude reaches 31.5~hPa, substantially higher than at 0°, confirming the amplifying effect of vertical orientation. Amplitude further increases under hypergravity, reaching 37~hPa. In contrast, under microgravity, pulsatility decreases sharply to 16~hPa, a value of the same order of magnitude as that measured at 0° under microgravity (Figure~\ref{fig:pic_pulsatility_boxplot}).

Thus, for the vertical orientation, ICP pulsatility is strongly amplified under normal gravity and hypergravity, whereas microgravity induces a pronounced attenuation of amplitude and a modification of waveform morphology, suggesting a reorganization of the intracranial system in the absence of gravitational loading.

\subsection{Influence of phantom orientation}

\paragraph{Comparison between 0° and 90° under normal gravity (1g).}

\begin{figure}[h]
    \centering
    \includegraphics[width=1.0\textwidth]{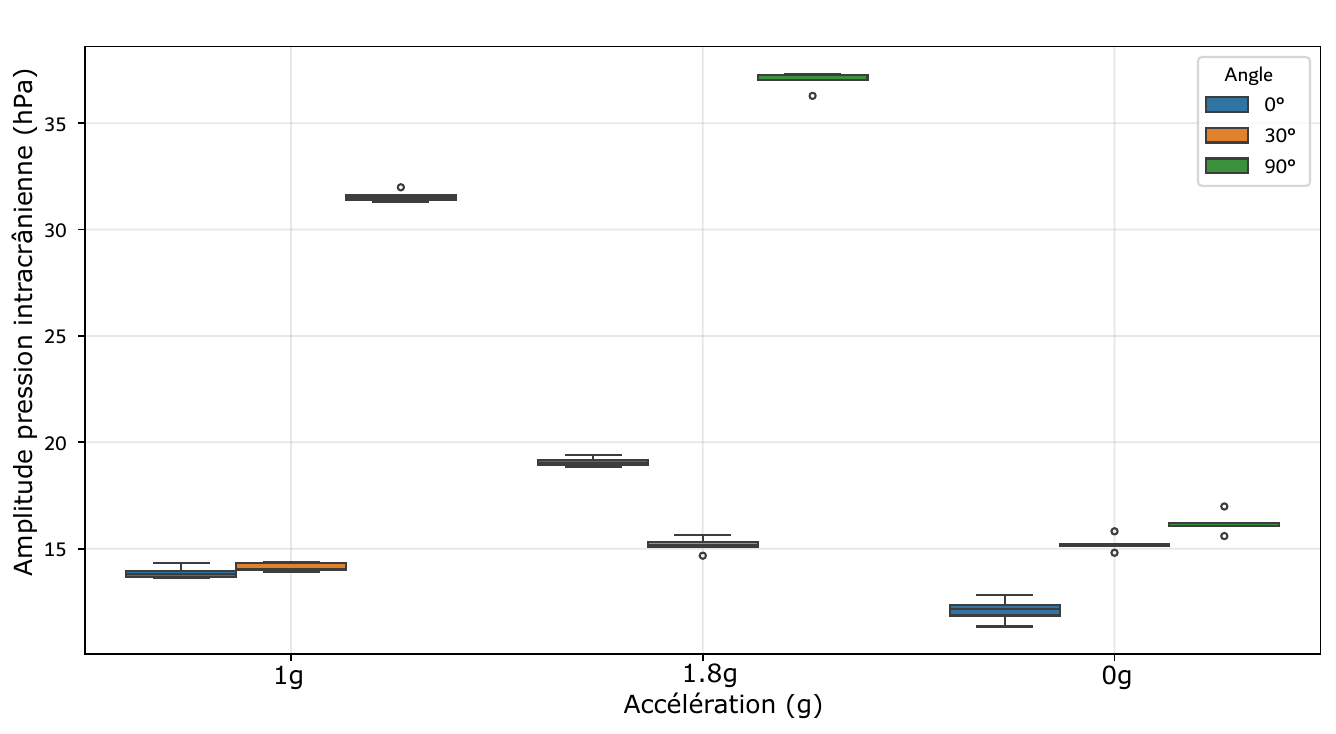}
    \caption{Peak-to-peak amplitude of intracranial pressure pulsatility as a function of gravitational conditions (1g, 1.8g, and 0g) and phantom orientation (0°, 30°, and 90°).}
    \label{fig:pic_pulsatility_boxplot}
\end{figure}

Under normal gravity, phantom orientation is a major determinant of intracranial pressure pulsatility. Transition from the horizontal position (0°) to the vertical position (90°) leads to a strong amplification of peak-to-peak amplitude, increasing from 13.9~hPa to 31.5~hPa, i.e., an increase of 17.6~hPa (Figure~\ref{fig:pic_pulsatility_boxplot}).

\begin{figure}[htbp]
\centering

% ===== Row 1 =====
\begin{subfigure}[t]{0.48\textwidth}
  \centering
  \includegraphics[
    angle=90,
    height=0.9\linewidth,
    keepaspectratio
  ]{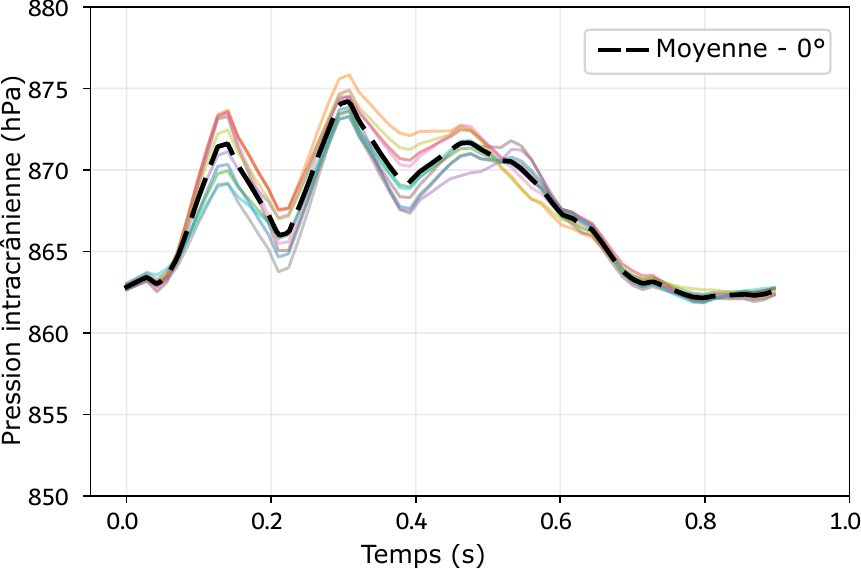}
   \vspace{2mm}
  \raisebox{30mm}{\rotatebox{90}{\small 0g — 0°}}
\end{subfigure}
\hfill
\begin{subfigure}[t]{0.48\textwidth}
  \centering
  \includegraphics[
    angle=90,
    height=0.9\linewidth,
    keepaspectratio
  ]{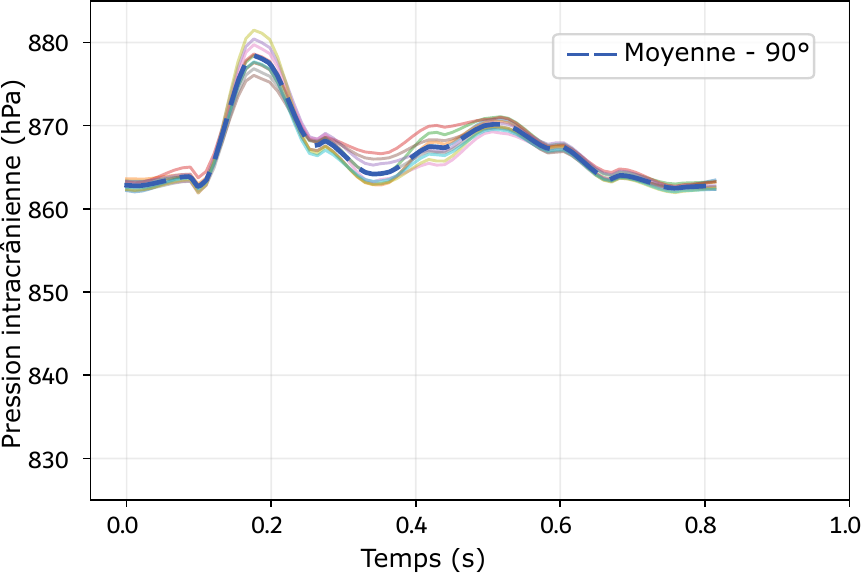}
  \vspace{2mm}
  \raisebox{30mm}{\rotatebox{90}{\small 0g — 90°}}
\end{subfigure}

\vspace{4mm}

% ===== Row 2 =====
\begin{subfigure}[t]{0.48\textwidth}
  \centering
  \includegraphics[
    angle=90,
    height=0.9\linewidth,
    keepaspectratio
  ]{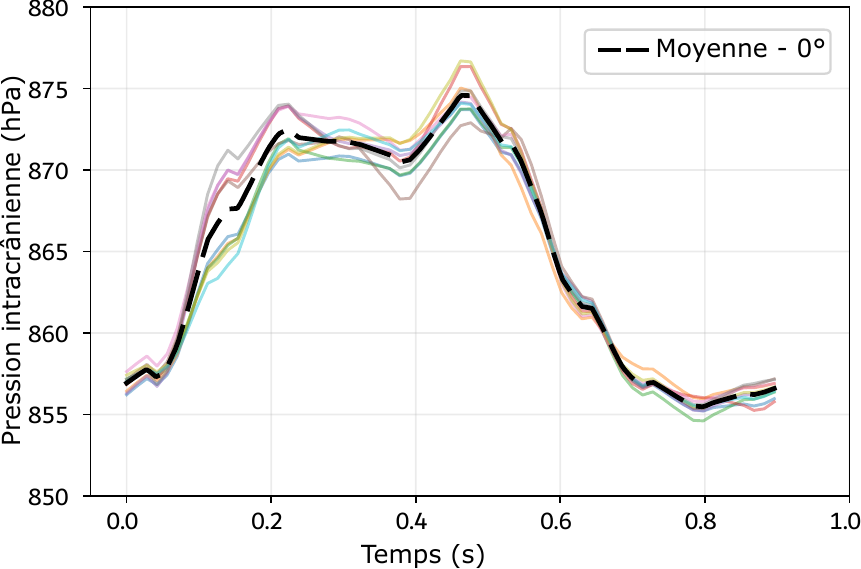}
  \vspace{2mm}
  \raisebox{30mm}{\rotatebox{90}{\small 2g — 0°}}
\end{subfigure}
\hfill
\begin{subfigure}[t]{0.48\textwidth}
  \centering
  \includegraphics[
    angle=90,
    height=0.9\linewidth,
    keepaspectratio
  ]{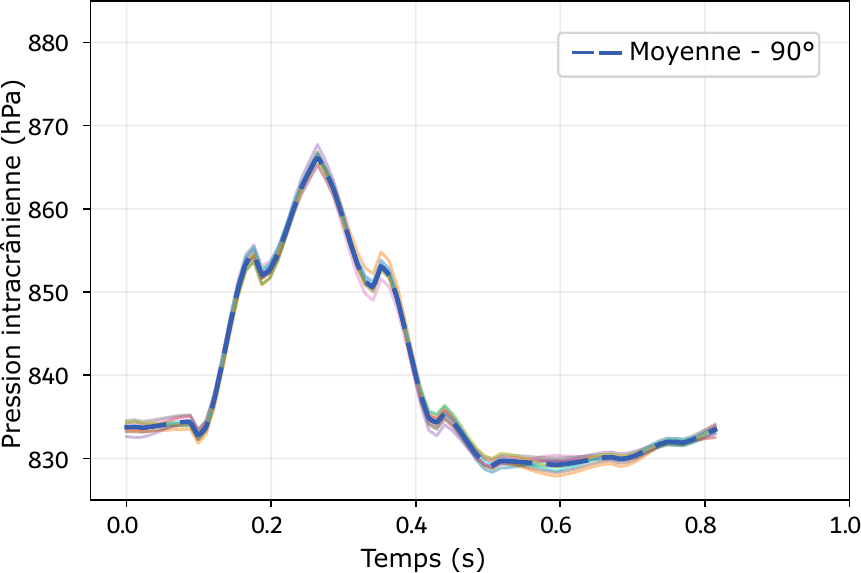}
  \vspace{2mm}
  \raisebox{30mm}{\rotatebox{90}{\small 2g — 90°}}
\end{subfigure}

\vspace{4mm}

% ===== Row 3 =====
\begin{subfigure}[t]{0.48\textwidth}
  \centering
  \includegraphics[
    angle=90,
    height=0.9\linewidth,
    keepaspectratio
  ]{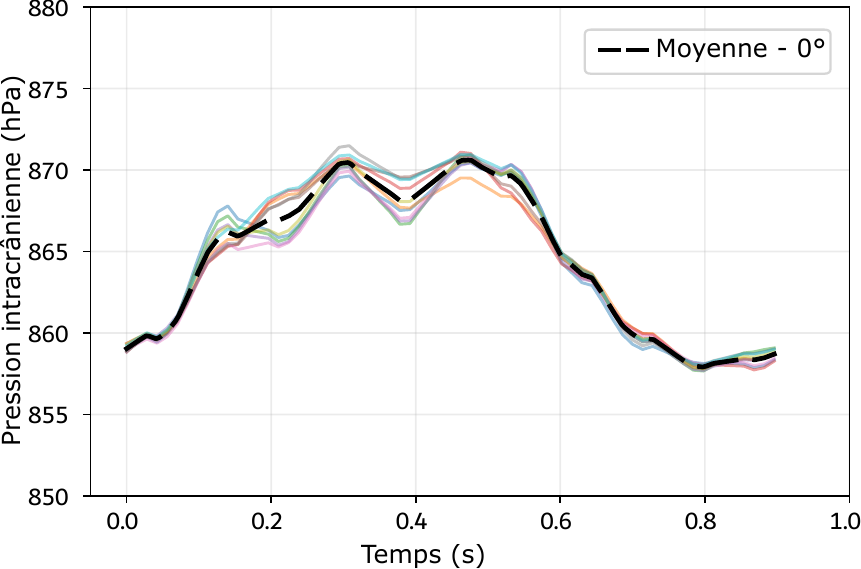}
  \vspace{2mm}
  \raisebox{30mm}{\rotatebox{90}{\small 1g — 0°}}
\end{subfigure}
\hfill
\begin{subfigure}[t]{0.48\textwidth}
  \centering
  \includegraphics[
    angle=90,
    height=0.9\linewidth,
    keepaspectratio
  ]{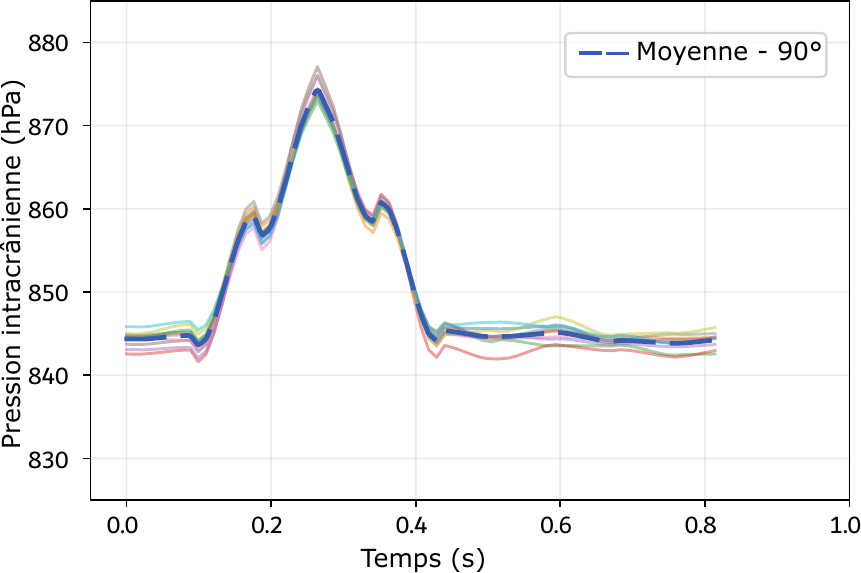}
  \vspace{2mm}
  \raisebox{30mm}{\rotatebox{90}{\small 1g — 90°}}
\end{subfigure}

\caption{Intracranial pressure waveforms normalized over a cardiac cycle for 0° and 90° orientations, and for gravity conditions 1g, 1.8g, and 0g. Colored curves represent ten successive cycles and the dashed curve indicates the temporal mean.\\
Pression intracrânienne: intracranial pressure; Temps: Time; Moyenne: Mean.}
\label{fig:pulsatile_pic_0_90}
\end{figure}

Beyond amplitude differences, orientation also modifies the temporal dynamics of ICP. Although three peaks are observed in both configurations, their relative organization differs: in the vertical position, the second peak becomes dominant and the peaks are concentrated over a shorter portion of the cardiac cycle, indicating stronger and more temporally compressed pulsatility (Figure~\ref{fig:pulsatile_pic_0_90}).

These results emphasize that even under normal gravity, changing system orientation alone is sufficient to profoundly alter intracranial pulsatile dynamics.

\paragraph{Comparison between 0° and 90° under microgravity (0g).}

Under microgravity, the effect of phantom orientation on intracranial pressure pulsatility is strongly attenuated compared with terrestrial gravity conditions. Peak-to-peak amplitude differs only slightly between the horizontal and vertical positions, with values of 12.3~hPa at 0° and 16~hPa at 90°, i.e., a limited difference of 3.7~hPa, far smaller than that observed at 1g (Figure~\ref{fig:pic_pulsatility_boxplot}).

While orientation dependence is reduced in terms of amplitude, differences persist in the temporal organization of pressure waves. In the horizontal position, the waveform retains a relatively marked multi-peak structure, whereas in the vertical position, pulsatile dynamics appear simplified, dominated by a more rounded main peak and an attenuated second peak (Figure~\ref{fig:pulsatile_pic_0_90}).

These observations suggest that under microgravity, the absence of gravitational loading limits the influence of orientation on ICP pulsatility amplitude, while still modulating waveform morphology.

\subsection{Intracranial and spinal pressure pulsatility}

\begin{figure}[htbp]
\centering

% ===== Row 1 =====
\begin{subfigure}[t]{0.48\textwidth}
  \centering
  \includegraphics[
    angle=90,
    height=0.9\linewidth,
    keepaspectratio
  ]{figures/results/Pic_0_0g.pdf}
   \vspace{2mm}
  \raisebox{30mm}{\rotatebox{90}{\small 0g — 0°}}
\end{subfigure}
\hfill
\begin{subfigure}[t]{0.48\textwidth}
  \centering
  \includegraphics[
    angle=90,
    height=0.9\linewidth,
    keepaspectratio
  ]{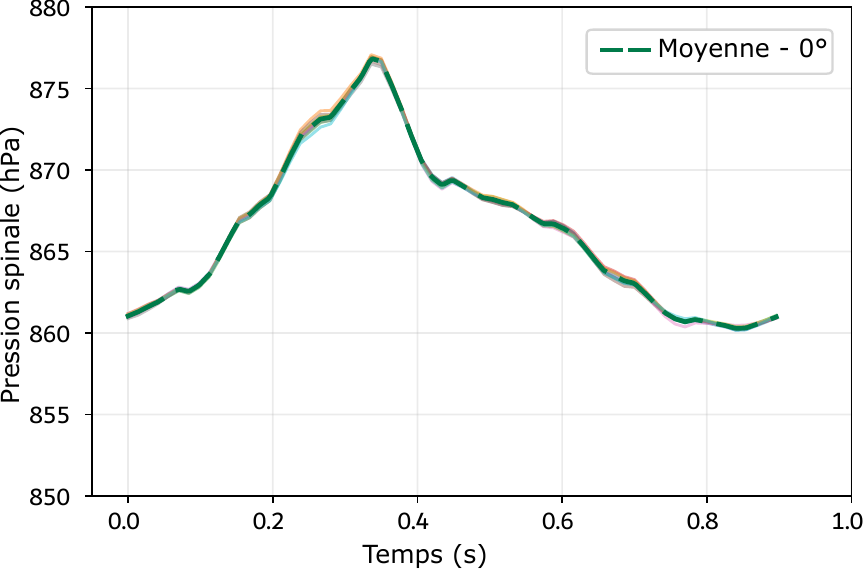}
  \vspace{2mm}
  \raisebox{30mm}{\rotatebox{90}{\small 0g — 0°}}
\end{subfigure}

\vspace{4mm}

% ===== Row 2 =====
\begin{subfigure}[t]{0.48\textwidth}
  \centering
  \includegraphics[
    angle=90,
    height=0.9\linewidth,
    keepaspectratio
  ]{figures/results/Pic_0_2g.pdf}
  \vspace{2mm}
  \raisebox{30mm}{\rotatebox{90}{\small 2g — 0°}}
\end{subfigure}
\hfill
\begin{subfigure}[t]{0.48\textwidth}
  \centering
  \includegraphics[
    angle=90,
    height=0.9\linewidth,
    keepaspectratio
  ]{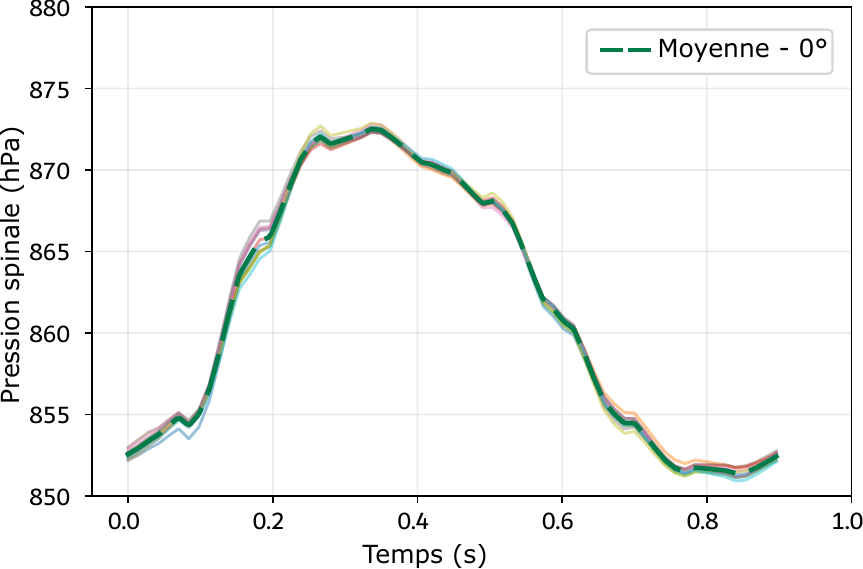}
  \vspace{2mm}
  \raisebox{30mm}{\rotatebox{90}{\small 2g — 0°}}
\end{subfigure}

\vspace{4mm}

% ===== Row 3 =====
\begin{subfigure}[t]{0.48\textwidth}
  \centering
  \includegraphics[
    angle=90,
    height=0.9\linewidth,
    keepaspectratio
  ]{figures/results/Pic_0_1g.pdf}
  \vspace{2mm}
  \raisebox{30mm}{\rotatebox{90}{\small 1g — 0°}}
\end{subfigure}
\hfill
\begin{subfigure}[t]{0.48\textwidth}
  \centering
  \includegraphics[
    angle=90,
    height=0.9\linewidth,
    keepaspectratio
  ]{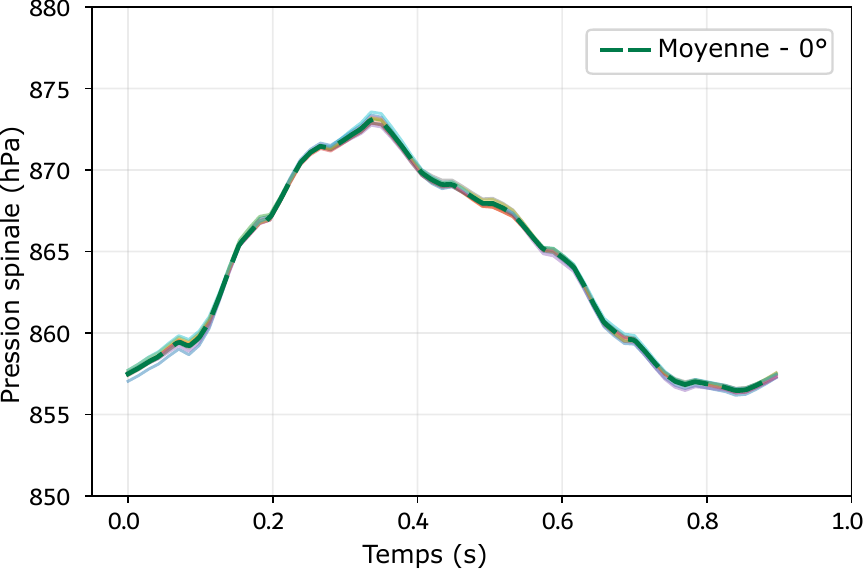}
  \vspace{2mm}
  \raisebox{30mm}{\rotatebox{90}{\small 1g — 0°}}
\end{subfigure}

\caption{Intracranial and spinal pressure waveforms normalized over a cardiac cycle in the horizontal position (0°) for gravity conditions 1g, 1.8g, and 0g. Colored curves represent ten successive cycles and the dashed curve indicates the temporal mean.\\
Pression intracrânienne: intracranial pressure; Pression spinale: Spinal pressure; Temps: Time; Moyenne: Mean.}
\label{fig:pulsatile_pic_pspinal_0}
\end{figure}

In the horizontal position (0°), ICP and spinal pressure waveforms exhibit distinct dynamics (Figure~\ref{fig:pulsatile_pic_pspinal_0}). ICP is characterized by a complex pulsatile structure with multiple successive peaks over the cardiac cycle.

In contrast, spinal pressure shows a smoother waveform, characterized by a progressive rise toward a single maximum followed by a decay phase. This difference in behavior is observed across all three gravity phases. Under microgravity, the spinal pressure peak appears slightly more pronounced, consistent with the modifications observed in ICP waveform morphology in this condition.

\newpage

\subsection{Qualitative comparison between human waveforms and the phantom model}

A comparative examination of the pressure waveforms measured in the phantom model and those classically reported in humans indicates that the phantom reproduces several fundamental features of physiological dynamics, while also exhibiting notable structural differences. This comparison was performed for the 1g condition and a horizontal phantom orientation (0°), corresponding to a configuration analogous to the supine position in humans.

\paragraph{Similarities.}

First, the phantom ICP waveform retains a polyphasic structure. Although transitions are less abrupt than in humans, three distinct peaks or inflection points can be identified over the cardiac cycle. This temporal organization reflects an attempt to reproduce the main physiological components commonly described in humans.

The morphology of spinal pressure constitutes another strong similarity between the phantom model and human physiology. In both cases, the spinal pressure curve is smoother and more rounded than the ICP waveform. It acts as a damping filter with respect to rapid ICP variations, suggesting a role in dissipation and redistribution of volume changes within the spinal compartment, both in humans and in the physical model.

\begin{figure}[htbp]
    \centering

    \begin{subfigure}[t]{0.7\textwidth}
        \centering
        \includegraphics[width=\linewidth]{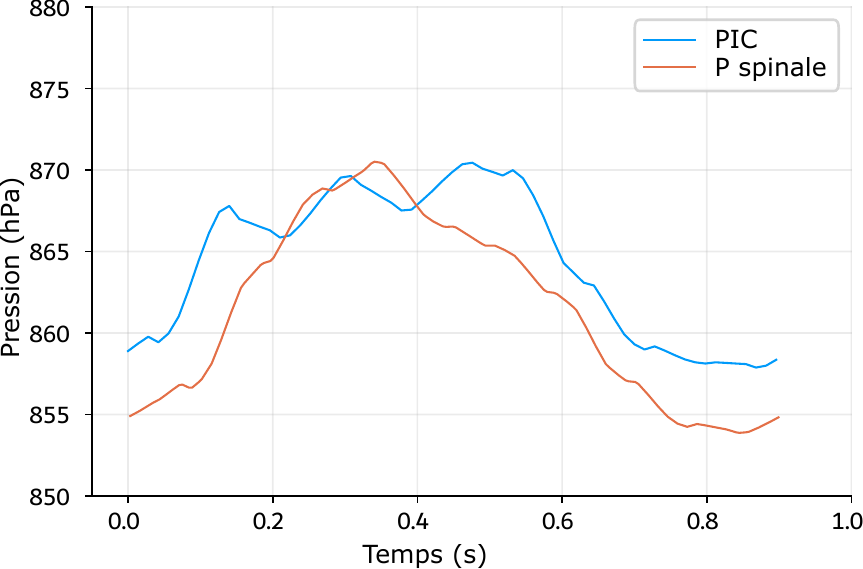}
        \caption{Phantom model --- normal gravity (1g), horizontal position (0°).}
        \label{fig:modele_fantome}
    \end{subfigure}

    \vspace{4mm}

    \begin{subfigure}[t]{0.7\textwidth}
        \centering
        \includegraphics[width=\linewidth]{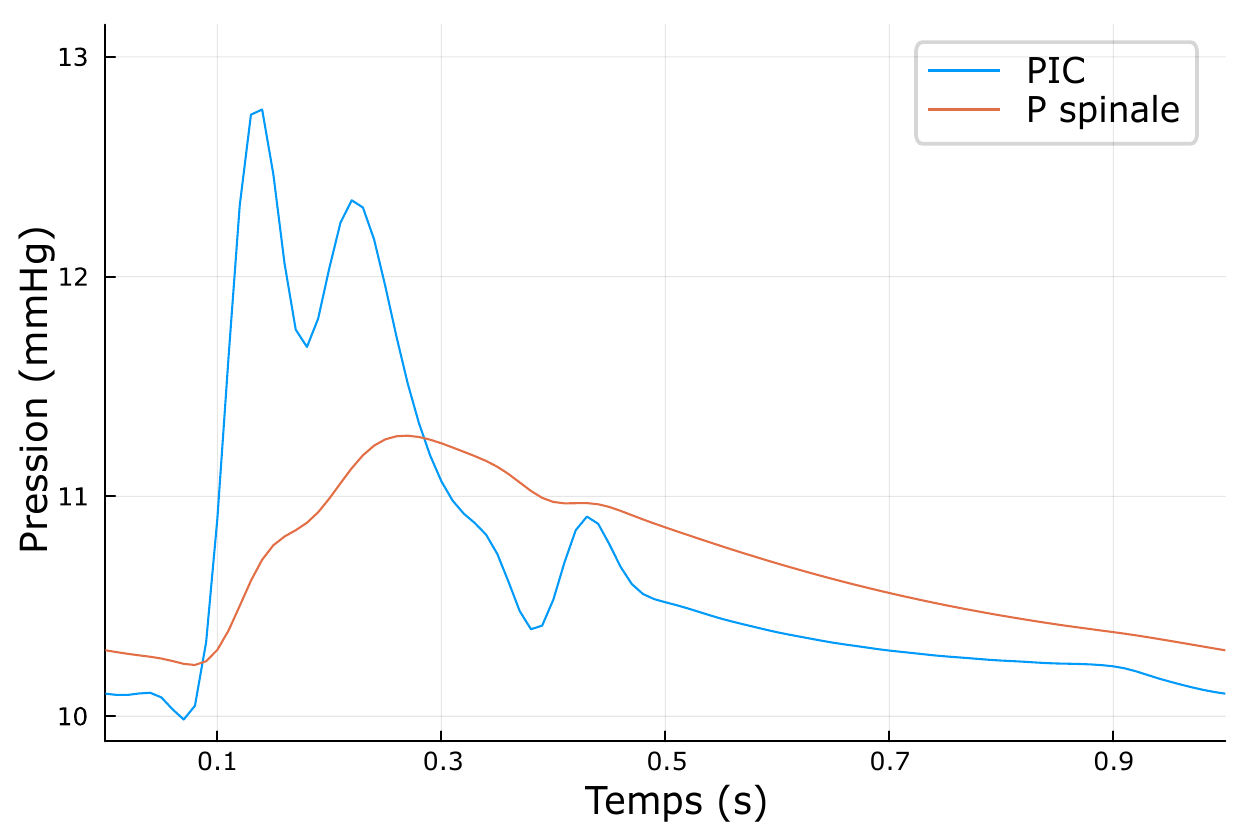}
        \caption{Human reference --- supine position.}
        \label{fig:modele_humain}
    \end{subfigure}

    \caption{Intracranial and spinal pressure waveforms measured in the phantom model and reported in humans, normalized over a cardiac cycle.\\
    Pression: pressure; PIC: ICP; P spinale: Spinal pressure; Temps: Time.}
    \label{fig:modele_fantome_humain}
\end{figure}

Finally, temporal consistency is observed between the two systems. Waveforms are cyclical and synchronized with a period of approximately one second, corresponding to one cardiac cycle. In both cases, the signal is characterized by an initial pressure rise followed by a slower relaxation phase (Figure~\ref{fig:modele_fantome_humain}).

\paragraph{Differences.}

Despite these qualitative similarities, differences emerge regarding signal amplitude and temporal concentration. In humans, ICP peaks are sharp and concentrated within a short time window, typically between 0.1 and 0.3~s of the cardiac cycle. In contrast, in the phantom model, although three peaks are present, they are more spread out in time and less pronounced, extending over a broader temporal range that can exceed 0.5~s. This difference reflects a more damped and less energetic dynamics in the experimental model.

Another difference concerns coupling between intracranial and spinal compartments. In humans, a strong contrast is typically observed between the pronounced ICP pulsatility and the relative stability of spinal pressure, whose amplitude remains lower. In the phantom model, although spinal pressure remains more rounded, its amplitude closely follows that of ICP. This behavior suggests more direct hydraulic communication between compartments or a less marked compliance differentiation than in the physiological human system.

Finally, relative pressure amplitudes differ between the model and humans. In the phantom model, ICP pulsatility amplitude reaches approximately 12~hPa (i.e., about 9~mmHg), whereas typical human ICP pulsatility is on the order of 3~mmHg. Moreover, while in humans spinal pressure amplitude is generally lower than ICP amplitude, the phantom model exhibits an inverse or comparable situation (Figure~\ref{fig:modele_fantome_humain}).

\chapter{Discussion}

\section{Evolution of mean pressure values}

Analysis of mean pressures highlights distinct behaviors of the intracranial and spinal compartments as a function of the gravity vector and phantom inclination. These results emphasize the central role of the hydrostatic component in regulating pressures within the experimental model, as well as the inherent limitations of a simplified physical system.

\subsection{Influence of inclination and hydrostatic pressure under gravity conditions}

Under terrestrial gravity and hypergravity, the opposite evolution of ICP and spinal pressure with phantom angle can be explained by variations in the hydrostatic fluid column.

When the phantom transitions from a horizontal to a vertical position, the intracranial sensor is located at a relatively higher position with respect to the fluid center of mass. Intracranial pressure therefore decreases with the vertical height $h$, in accordance with the hydrostatic law:
\[
\Delta P = \rho \cdot g \cdot h
\]
where $\rho$ is the fluid density, $g$ is the gravitational acceleration, and $h$ is the height of the fluid column. This relation accounts for the progressive decrease in ICP observed as inclination increases. Under hypergravity, the increase in acceleration $g$ amplifies this effect, which explains the larger ICP drop measured experimentally (approximately 25~hPa versus 15~hPa at 1g).

Conversely, the spinal compartment is located at a relatively lower position when the phantom is inclined. The increase in the fluid column height above the spinal sensor leads to an increase in the measured pressure. This opposite evolution of intracranial and spinal pressures therefore constitutes a direct signature of hydrostatic effects within the model.

A specific feature is nevertheless observed under normal gravity between the 30° and 90° positions, where spinal pressure shows an almost stagnant behavior (866~hPa versus 866.3~hPa). This lack of a marked increase may reflect a limitation of the effective compliance of the phantom spinal compartment. Another plausible explanation is a measurement bias related to sensor positioning in the vertical configuration. A slight sensor drift or an edge effect linked to the phantom geometry at 90° could attenuate the expected variation, unlike in hypergravity where the hydrostatic effect is amplified.

\subsection{Mean pressure behavior under microgravity}

Under microgravity, the disappearance of the hydrostatic gradient leads to pressure homogenization throughout the phantom. Experimental results show a mean ICP that is essentially identical regardless of phantom orientation, with values close to 867~hPa. This behavior is consistent with the absence of fluid-column weight and confirms that, under these conditions, pressure is no longer governed by the relative height of the sensor.

However, ICP values measured under microgravity appear higher than those observed at 1g for inclined configurations. This result contrasts with human observations reported by Lawley \textit{et al.}, 2017, where ICP measured under microgravity is generally lower than that measured in the supine position on Earth \cite{Lawley2017}.

This discrepancy can be explained by the inert nature of the phantom model. In humans, ICP under microgravity is regulated by complex physiological mechanisms, including body fluid redistribution and a marked cephalad shift of blood, together with venous adaptations. In contrast, the phantom model includes no active regulation mechanism.

In this context, the ICP increase observed under microgravity may result from several factors. The absence of gravity-assisted drainage may increase the simulated venous return pressure within the circuit, which directly impacts ICP. In addition, ambient cabin pressure variations during parabolic flights, combined with the hydraulic circuit layout and pump operation, may elevate the baseline system pressure. In the absence of an opposing hydrostatic gradient, this baseline elevation is fully transmitted to the measured intracranial pressure.

\section{Pulsatile dynamics and pressure waveforms}

Analysis of pressure waveform evolution as a function of phantom inclination and gravitational conditions reveals structural modifications in the pulsatile dynamics of the system. These variations primarily reflect changes in the effective compliance of the phantom, i.e., its ability to absorb and redistribute the pulsatile volume imposed by the hydraulic excitation.

\subsection{Effect of phantom orientation: role of spinal compliance}

Under normal gravity, transitioning from the horizontal to the vertical position leads to a marked increase in ICP pulsatility amplitude, from 13.9~hPa to 31.5~hPa. This increase is accompanied by a notable change in waveform morphology, with a stronger dominance of the second peak.

This behavior can be interpreted through the role of the phantom's spinal compliance. In the vertical position, hydrostatic pressure increases in the lower part of the system, corresponding to the spinal compartment. This pressure increase stretches the deformable membrane in the spinal section, making it mechanically stiffer.

The direct consequence of this stiffening is a reduction in spinal compliance. The spinal compartment, which plays a damping role in the horizontal position, therefore loses part of its ability to absorb the volume variations imposed by the pump. The pressure wave is then transmitted more abruptly, and with larger amplitude, toward the intracranial compartment, which is structurally more rigid. This redistribution explains both the global increase in pulsatile amplitude and the predominance of the second peak, which is often associated with wave reflections and the compliance properties of the system.

\subsection{Comparison between intracranial and spinal pressure: smoothing effect}

The morphological difference observed between ICP waveforms and spinal pressure waveforms is a key element for understanding model behavior. While ICP exhibits a complex pulsatile structure with several successive peaks, spinal pressure is characterized by a smoother waveform dominated by a single maximum over the cardiac cycle.

This observation confirms that the phantom spinal compartment acts as a low-pass filter with respect to pressure variations. The higher compliance of the spinal section enables absorption of high-frequency components of the pressure wave generated by the pump, leaving mainly a gradual overall pressure increase. This behavior is qualitatively analogous to the Windkessel effect observed in the human vascular system, where large elastic arteries smooth the pulsatile flow generated by the heart.

When spinal compliance decreases, notably in the vertical position under the effect of hydrostatic pressure, this filtering capacity is reduced. ICP pulsatile dynamics then become more pronounced, with increased amplitude and a modified temporal distribution of peaks.

\subsection{Pulsatile dynamics under microgravity: system relaxation}

Under microgravity, pulsatile dynamics exhibit an intermediate behavior. ICP pulsatility amplitude decreases compared with the vertical configuration under normal gravity, but waveforms appear more pronounced or more ``nervous'' than in some 1g configurations.

A plausible hypothesis is based on mechanical relaxation of the phantom in the absence of the water-column weight. Under microgravity, deformable membranes are no longer stretched by hydrostatic pressure, regardless of system orientation. The phantom therefore exhibits a more homogeneous compliance between intracranial and spinal compartments, without gravity-induced pre-tension.

This relaxation could explain why waveforms measured under microgravity at 90° become closer to those observed at 0°, since the pressure gradient responsible for spinal-compartment stiffening disappears. The ``more pulsatile relaxation'' observed at 0g suggests that the pressure wave propagates in a medium with more compliant walls, allowing freer expression of hydraulic resonance phenomena intrinsic to the experimental circuit.

\subsection{Model limitations and lack of direct human data}

It should be emphasized that, in humans, the evolution of ICP waveform morphology under microgravity remains poorly documented due to the invasive nature of the required measurements. Direct comparison between phantom waveforms and human physiology in microgravity is therefore limited.

In the phantom model, the increase in pulsatility observed under hypergravity indicates that the mechanical stress applied to the system reduces its compensatory reserve capacity. Conversely, the marked attenuation of the orientation effect under microgravity, with an amplitude difference limited to approximately 3.7~hPa between 0° and 90°, indicates that posture becomes a major determinant of pulsatility only when a gravity vector is present to modify baseline pressure distribution within the system.

\section{Comparison with human physiology and model limitations}

The phantom model provides an original experimental approach to study intracranial and spinal pressures in the presence of a variable gravity vector. To our knowledge, few physical models simultaneously reproduce a cardiovascular-like pulsatile excitation while incorporating the effects of gravity and posture on cranial and spinal compartments. The measured waveforms suggest that pulsations observed in the phantom primarily arise from reverberation and reflection of pressure waves within a rigid cranial structure coupled to a deformable spinal compartment.

Although the phantom successfully reproduces the hydrostatic gradient and the influence of posture on overall system compliance, several simplifications limit direct translation of the results to human physiology.

\subsection{Lack of active regulation of pulsatile flow}

A major limitation of the model lies in the constancy of pump parameters across gravity phases. In humans, transitions to hypergravity or microgravity induce complex reflex cardiovascular responses, such as baroreflex activation, changes in heart rate, and adaptations in stroke volume.

In this study, the flow imposed by the pump was slightly modified from one phase to another, but only to a limited extent. In real physiological conditions, the interaction between the heart and gravity modifies the energy of the incident arterial pressure wave, which directly influences the amplitude and temporal distribution of ICP peaks. The absence of such active regulation in the model therefore represents an important simplification.

\subsection{Role of venous and thoracic pressures}

The experimental model primarily focuses on the pulsatile input simulated by the pump and on hydrostatic effects linked to the fluid column. However, many studies emphasize that ICP is sensitive to central venous pressure and intrathoracic pressure.

In humans, respiratory-related thoracic pressure variations, as well as ambient pressure changes, strongly influence cerebral venous drainage. During parabolic flights, these effects are exacerbated by rapid acceleration transitions. These mechanisms are not accounted for in the phantom model, which includes neither a thoracic compartment nor respiratory dynamics.

In addition, in humans, jugular veins behave as a \textit{Starling resistor}: in the upright position, partial collapse introduces a hydraulic resistance that protects ICP from excessively abrupt variations. In the phantom model, venous conduits are rigid or semi-rigid and do not reproduce this collapse mechanism. This absence likely explains the more direct hydraulic communication observed between intracranial and spinal compartments, as well as the higher pressure amplitudes measured in the phantom (approximately 9~mmHg for ICP, versus typically 3~mmHg in humans).

\subsection{Compliance differentiation between compartments}

Finally, the divergence observed between relative amplitudes of ICP and spinal pressure highlights the need to refine compliance differentiation between model compartments. In humans, the spinal dural sac provides a substantial expansion reserve, playing a key role in damping intracranial pulsations.

In the phantom model, the similarity between ICP and spinal pressure amplitudes indicates that pulsatile energy is transmitted with limited losses between compartments. This behavior suggests that the hydraulic interface between cranial and spinal compartments is likely oversimplified compared with the complexity of the human subarachnoid space. Improving compliance differentiation is therefore an essential perspective to increase the biomechanical realism of the model.

\chapter{Conclusion and perspectives}

The MODÈFONE project enabled the development and operation of an original phantom model of the cerebrospinal system, integrating both a cardiovascular-like pulsatile excitation and the explicit consideration of the gravity vector. This experimental device made it possible to jointly study intracranial and spinal pressures under variable gravity conditions during parabolic flights, thereby providing an experimental framework that is rarely accessible.

The results demonstrate that the model coherently reproduces hydrostatic effects related to posture under normal gravity and hypergravity, with opposite evolutions of intracranial and spinal pressures as a function of inclination. Under microgravity, the homogenization of mean pressures and the attenuation of posture effects confirm the central role of the hydrostatic gradient in the passive regulation of pressures. Analysis of pulsatile dynamics highlights a strong dependence of waveform amplitude and morphology on both orientation and gravity, emphasizing the key role of spinal compliance in damping pressure variations.

Qualitative comparison with human physiology shows that the phantom is capable of reproducing several fundamental characteristics of physiological waveforms, notably their polyphasic nature and the damping role of the spinal compartment. However, discrepancies remain, particularly regarding pressure amplitudes and the coupling between intracranial and spinal compartments, reflecting the simplifications inherent to an inert physical model lacking active regulation.

In this context, several perspectives for improvement can be considered to enhance the biomechanical realism and scientific scope of the model. A first development would consist in systematically evaluating the influence of pump parameters (ejection fraction, input signal shape) on intracranial pressure waveform morphology, in order to better understand the relationship between hydraulic excitation and the pulsatile response of the system.

The integration of flow and pressure sensors at the inlet of the hydraulic circuit would also represent a major advancement. Such measurements would enable finer correlation between pressures measured within the phantom and system input parameters, and would allow more precise identification of pressure-wave transmission and reflection mechanisms.

Furthermore, the development of a more elaborate venous component represents an essential perspective. Introducing deformable venous conduits capable of reproducing collapse phenomena, particularly at the level of the jugular veins, would improve the representation of physiological adaptations associated with gravity changes and enhance modeling of cerebral venous drainage.

Finally, exploring different materials for the deformable parts of the phantom constitutes a key avenue to adjust the distribution of compliance between cranial and spinal compartments. Improved differentiation of the mechanical properties of these compartments would allow the model’s pulsatile dynamics to more closely approach those observed in humans.

In conclusion, the MODÈFONE model represents a robust first step toward the development of gravity-sensitive experimental models of the cerebrospinal system. It provides a relevant basis for exploring the physical mechanisms underlying intracranial and spinal pressure variations under variable gravity conditions, while opening the way for future developments aimed at progressively integrating more complex physiological mechanisms.

\paragraph{Acknowledgements}

This work could not have been carried out without the support of numerous institutions and partners. We would like to thank IMT Mines Alès and the University of Picardie Jules Verne for their institutional support, as well as the CHIMERE (UR~7516) and SAINBIOSE (INSERM~1059) laboratories for their scientific supervision and guidance throughout the project.

We also extend our sincere thanks to our sponsors—the IMT Foundation, the Foundation of the University of Picardie Jules Verne, SEMAXONE, and ArianePlast—for their essential financial and material support in the realization of the MODÈFONE project.

Finally, we would like to particularly thank all individuals and teams who contributed to this project and without whom none of this would have been possible, notably CNES and Novespace, for their trust, logistical support, and access to parabolic flights.

% Annexes
\appendix
\chapter{Appendix}

\section{Photographs of the experimental setup}

\begin{figure}[!htbp]
    \centering
    \begin{subfigure}[t]{0.7\textwidth}
        \centering
        \includegraphics[width=\linewidth]{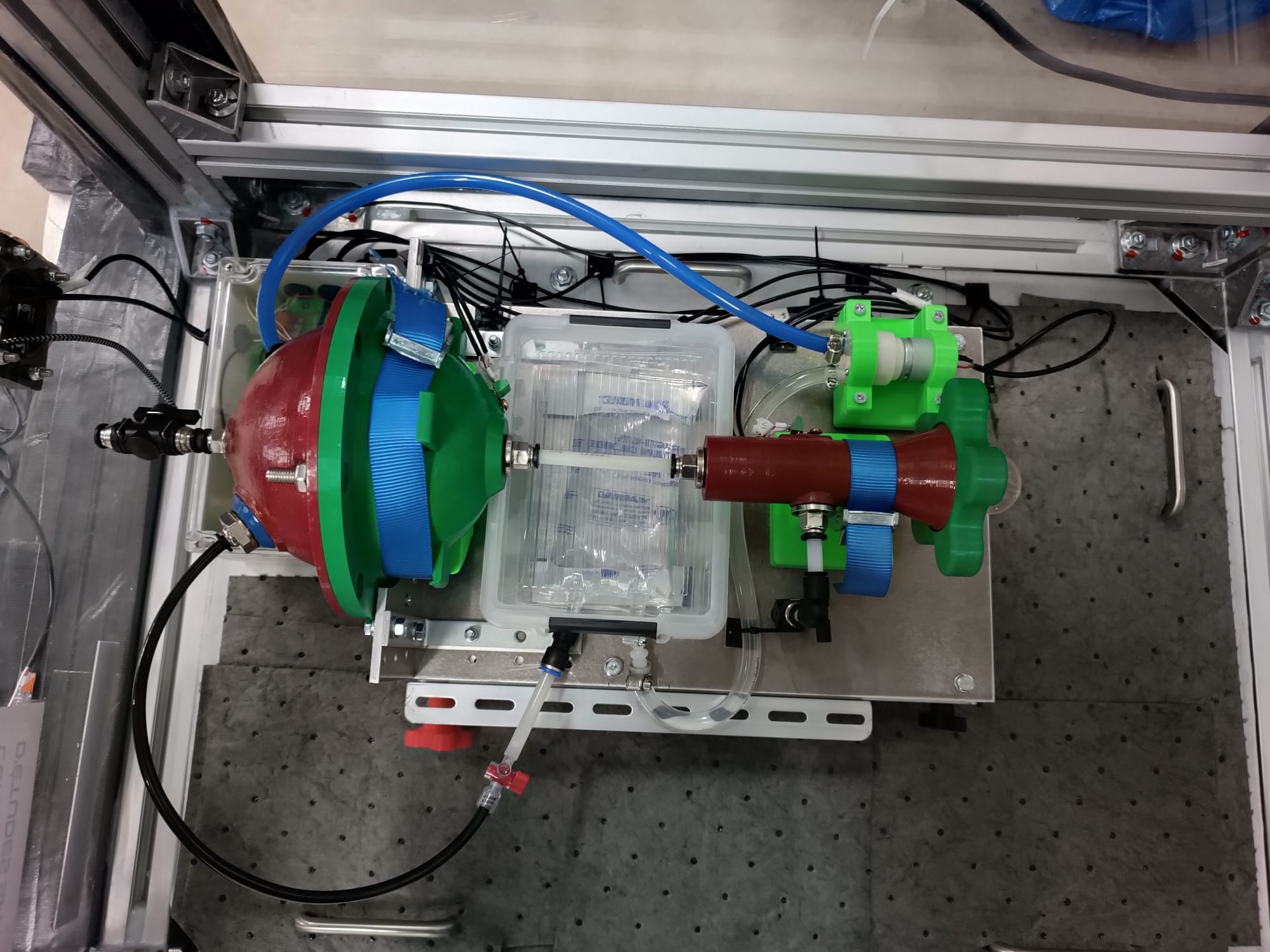}
    \end{subfigure}

    \vspace{2mm}

    \begin{subfigure}[t]{0.7\textwidth}
        \centering
        \includegraphics[width=\linewidth]{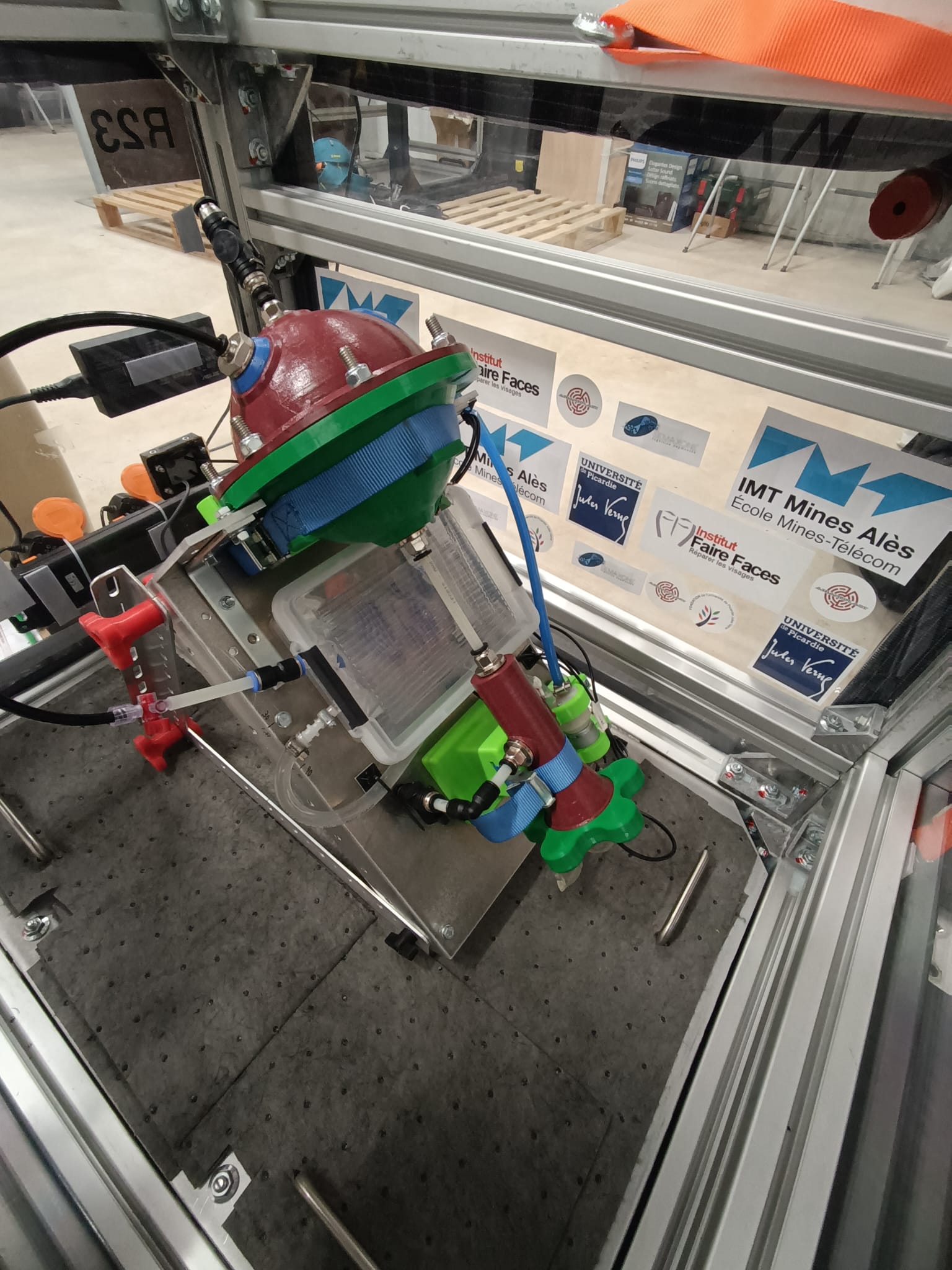}
    \end{subfigure}

    \caption{Phantom model: experimental views (1/2).}
    \label{fig:fantome_1}
\end{figure}

\clearpage

\begin{figure}[!htbp]
    \centering
    \includegraphics[width=0.6\textwidth]{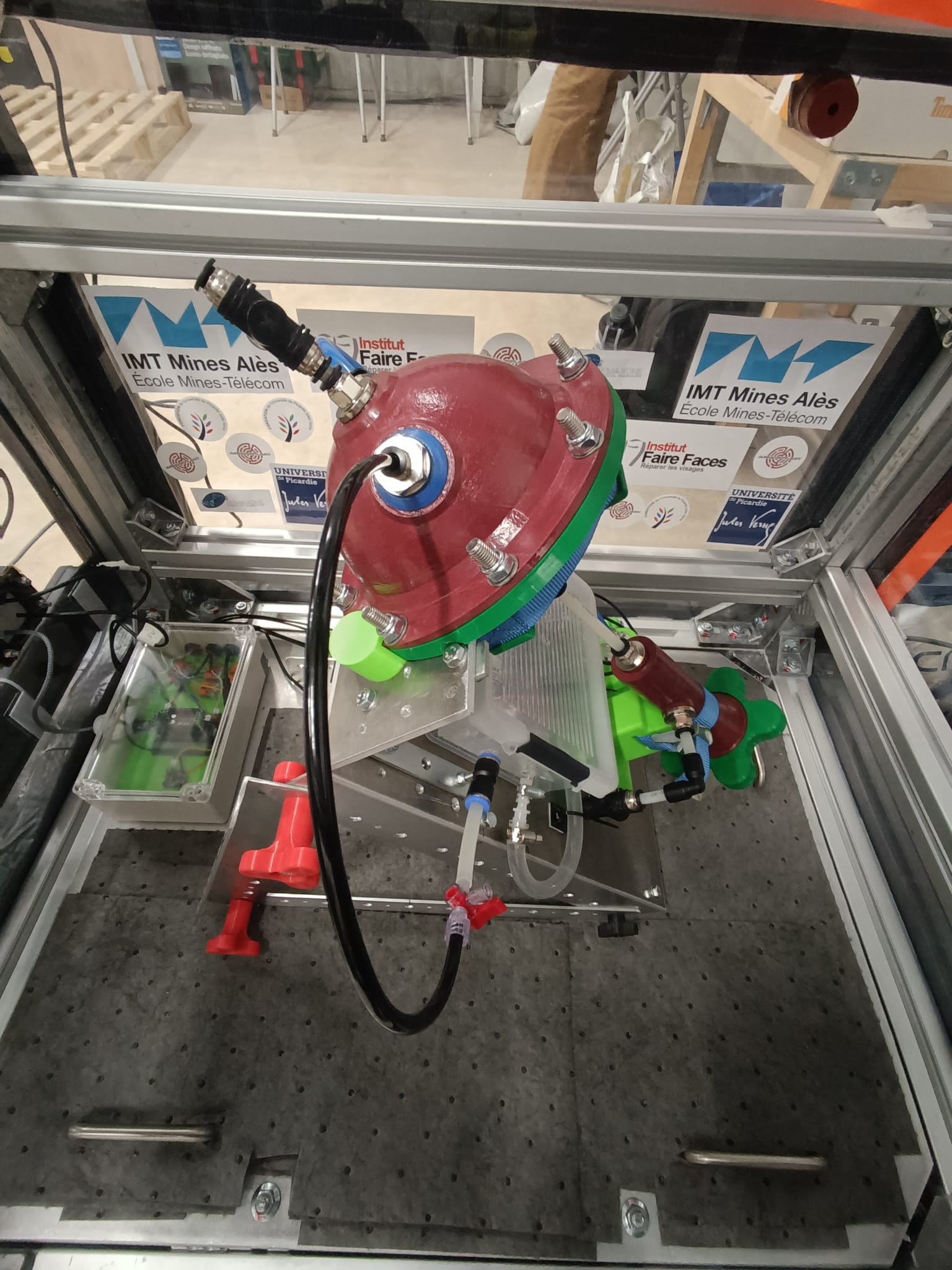}
    \caption{Phantom model: complementary experimental view (2/2).}
    \label{fig:fantome_2}
\end{figure}

% Bibliographie
\bibliographystyle{plain}
\bibliography{references}

\end{document}